\documentclass[twocolumn,showpacs,preprintnumbers,amsmath,amssymb,
superscriptaddress,nofootinbib]{revtex4}
\usepackage{graphicx}

\newcommand{\eps}{\varepsilon}
\newcommand{\tens}[1]{{\boldsymbol{#1}}}

\newcommand{\pa}{{\tens{\partial}}}

\newcommand{\cv}[1]{{\tens{\partial}}_{#1}}

\newcommand{\ph}{\varphi}

\newcommand{\be}{\begin{equation}}
\newcommand{\ee}{\end{equation}}
\newcommand{\ba}{\begin{eqnarray}}
\newcommand{\ea}{\end{eqnarray}}

\newcommand{\hhh}{\, ,\hspace{0.2cm}}

\newcommand{\hook}{\raisebox{-0.35ex}{\makebox[0.6em][r]
{\scriptsize $-$}}\hspace{-0.15em}\raisebox{0.25ex}{\makebox[0.4em][l]{\tiny
 $|$}}}
\newcommand{\eq}[1]{(\ref{#1})}

\newcommand{\n}[1]{\label{#1}}
\newcommand{\bs}[1]{{\boldsymbol{#1}}}

\newcommand{\bsi}[2]{{}^{\tiny #1}\!{\boldsymbol{#2}}}

\newcommand{\srh}{\frac1{\sqrt2}}
\newcommand{\mom}{\tens{l}_{+\!}}                            % the momentum
\newcommand{\pfr}{{}^{\scriptscriptstyle\parallel}\mspace{-1.5mu}}     % parallelly transported frame
\newcommand{\bfr}{{}^{\scriptscriptstyle\mathrm{B}}\mspace{-1.5mu}}    % boosted frame
\newcommand{\nfr}{{}^{\scriptscriptstyle\mathrm{N}}\mspace{-1.5mu}}    % null-rotated frame

\newcommand{\kv}{{\tens{n}}}                           % real null vector k
\newcommand{\lv}{{\tens{l}}}                           % real null vector l
\newcommand{\muv}[1]{{\tens{m}_{\hat{#1}}}}                  % null vector in $\mu$th 2-plane                     %:ex: \muv{\mu}
\newcommand{\mbv}[1]{{\bar{\tens{m}}_{\hat{#1}}}}            % barred null vector in $\mu$th 2-plane              %:ex: \mbv{\mu}
\begin{document}

\title{Parallel-propagated frame along null geodesics in higher-dimensional black hole spacetimes}

\author{David Kubiz\v n\'ak}

\email{dk317@cam.ac.uk}

\affiliation{Theoretical Physics Institute, University of Alberta, Edmonton,
Alberta, Canada T6G 2G7}

\affiliation{DAMTP, University of Cambridge, Wilberforce Road, Cambridge CB3 0WA, UK}

\author{Valeri P. Frolov}

\email{frolov@phys.ualberta.ca}

\affiliation{Theoretical Physics Institute, University of Alberta, Edmonton,
Alberta, Canada T6G 2G7}

\author{Pavel Krtou\v{s}}

\email{Pavel.Krtous@utf.mff.cuni.cz}

\affiliation{Institute of Theoretical Physics, Faculty of Mathematics and Physics, Charles University in Prague,
V~Hole\v{s}ovi\v{c}k\'ach 2, Prague, Czech Republic}

\author{Patrick Connell}

\email{pconnell@phys.ualberta.ca}

\affiliation{Theoretical Physics Institute, University of Alberta, Edmonton,
Alberta, Canada T6G 2G7}

\date{October 31, 2008}  

\begin{abstract} In [arXiv:0803.3259] the equations describing the parallel
transport of  orthonormal frames along timelike (spacelike) geodesics in a spacetime
admitting a non-degenerate principal conformal Killing--Yano 2-form
$\bs{h}$ were solved. 
The construction employed is based on studying the Darboux subspaces of the 2-form $\tens{F}$
obtained as a projection of $\bs{h}$ along the geodesic trajectory.  
In this paper we demonstrate that, although slightly modified,  a similar construction is possible 
also in the case of null geodesics.
In particular, we explicitly construct the parallel-transported frames along null geodesics in $D=4,5,6$ Kerr-NUT-(A)dS spacetimes. 
We further discuss the parallel transport along principal null directions in these spacetimes. Such directions coincide with the eigenvectors of the principal conformal Killing--Yano tensor.
Finally, we show how to obtain a parallel-transported frame along null geodesics in the background of the 4D Pleba\'nski--Demia\'nski metric which admits only a conformal
generalization of the Killing--Yano tensor.
\end{abstract}

\pacs{04.50.-h, 02.40.-k, 04.50.Gh, 04.20.Jb \hfill  Alberta-Thy-08-08, DAMTP-2008-97}

\maketitle

\section{Introduction} 

Solving the parallel transport equations along null geodesics
in a four-dimensional spacetime 
is a well known problem with many physical applications. For example,
in a geometric optics approximation linearly polarized photons and gravitons propagate 
along null geodesics while the corresponding polarization
vectors are parallel-transported along the worldline \cite{MTW} (see also Appendix A). This
property was used to study the polarized radiation from black
holes (see, e.g., \cite{StarkConnors:1977, ConnorsStark:1977, ConnorsEtal:1980} and references
therein).  The parallel-propagated frames are convenient for
studying the form and shape of a thin `pencil of light' propagating in
an external gravitational field. In the derivation of the equations
for optical scalars the parallel-propagated frames play an
important technical role (see, e.g., \cite{Pirani:1965, Frolov:1977}). Another problem where such frames are useful is a so-called peeling-off property of the gravitational radiation in an asymptotically flat spacetime (see, e.g., \cite{Sachs:1961, Sachs:1962, NewmanPenrose:1962, Penrose:1965, KrtousPodolsky:2004b} and references therein). 

Recently, models of gravity with  extra dimensions attracted a lot of
attention. In order to re-derive many of the important 
results of the four-dimensional optics in a curved spacetime in the
higher-dimensional case it is necessary to consider first a problem of
parallel transport along null geodesics. This is the main purpose of this
paper.

We focus our attention on a special class of higher-dimensional
spacetimes which admit a so-called {\em principal conformal Killing--Yano}
(PCKY) tensor \cite{FrolovKubiznak:2007, KrtousEtal:2007jhep} (see also \cite{Frolov:2008, FrolovKubiznak:2008, Kubiznak:phd} for reviews).
The most general metric element admitting such a tensor, the {\em canonical} (off-shell) metric, was studied in \cite{HouriEtal:2007, KrtousEtal:2008}. Similar to 4D, when the vacuum Einstein equations with the cosmological constant are imposed the canonical metric becomes the Kerr-NUT-(A)dS spacetime \cite{ChenEtal:2006cqg}.

Recently, a parallel-propagated frame along timelike (spacelike) geodesics in the canonical spacetime was constructed \cite{ConnellEtal:2008}. 
As it often happens, a limit when the
velocity of the particle motion tends to the speed of light is
singular, so that, the problem of parallel transport along null
geodesics requires a special treatment. We deal with this  problem
in the present paper. The obtained results  generalize the 4D
results \cite{Marck:1983kerr, Marck:1983null, KamranMarck:1986}. 

Consider a spacetime  with a PCKY tensor $\tens{h}$, that is, a closed
rank-2 non-degenerate conformal Killing--Yano tensor
\cite{Tachibana:1969, Kashiwada:1968}.  In such a spacetime the
geodesic motion is completely integrable \cite{PageEtal:2007,
KrtousEtal:2007prd, HouriEtal:2008a}. Let us concentrate on a {\em generic} null ray and denote its velocity
vector $\tens{l}$. In our construction, starting with $\tens{l}$ we
first generate two additional parallel-propagated vectors, one of
which, say $\tens{n}$, is `external' to the null plane of $\tens{l}$
and can be made null. This means that the tangent space $T$ at each
point of $\tens{l}$ splits into a 2-dimensional parallel-propagated
subspace $U$ spread by $\{\tens{l},\tens{n}\}$ and a
$(D-2)$-dimensional parallel-propagated subspace $V$ orthogonal to
$U$; $T=U\oplus V$.

The construction of the parallel-propagated frame is now similar to
the timelike case. We consider the Darboux problem for the 2-form
$\tens{F}$  obtained as a projection of the PCKY tensor $\bs{h}$  to
a subspace $V$.  Such a 2-form is automatically parallel-transported.
In particular,  each of the Darboux subspaces of $\tens{F}$ is
independently parallel-transported. The zeroth
value Darboux subspace is 3(4)-dimensional  in an
odd(even)-dimensional spacetime. The parallel-propagated vectors
spanning it  can be explicitly constructed and we denote them 
$\{\tens{l},\tens{n},\tens{m}\}$
($\{\tens{l},\tens{n},\tens{m},\tens{z}\}$). The Darboux subspaces,
for non-zero eigenvalues of $\tens{F}$ are
2-dimensional. One can obtain the parallel-propagated  vectors
spanning them by employing 2D (affine parameter)-dependent rotations.

For certain special geodesics the construction of a parallel-propagated frame
has to be accordingly modified. 
Similar to the timelike (spacelike) case, the degeneracies occur 
for special geodesics for which the Darboux subspaces of $\tens{F}$ become 
more-dimensional. In these cases the frame can be obtained by ``more involved'' 
(affine parameter)-dependent orthogonal transformations (see \cite{ConnellEtal:2008} for more details).
An important example of different degeneracy is the 
case of parallel transport along principal null directions---discussed in Section~VII.  

The paper is organized as follows:
In Section~II we introduce the basic notations used in the paper and review  the definition of the PCKY tensor. In Section~III a generic construction of the parallel-propagated frames along null geodesics is outlined. 
Canonical metric element of a spacetime admitting the PCKY tensor
and its basic properties are discussed in Section~IV. In Section~V 
the explicit form of the canonical metric is employed to concretize the generic construction. Examples of solutions in four, five, and six dimensional
spacetimes are presented in Section~VI.  The parallel transport along principal null directions is described in Section~VII. Section~VIII is devoted to 
discussion of obtained results and conclusions. More technical details, connected
with the geometric optics in higher-dimensional spacetimes,  principal null directions, and the parallel-transport in the Pleba\'nski-Demia\'nski metric are presented in
Appendices A, B, and C, respectively.

\section{Basic notions and notations}

In what follows we use the notations of \cite{FrolovKubiznak:2008, ConnellEtal:2008}. We consider a $D$-dimensional spacetime $M^D$,
equipped with the metric 
\be\label{1}
\tens{g}=g_{ab}\tens{d}x^{a}\tens{d}x^{b}\, .
\ee
To  treat both cases of even and odd  dimensions simultaneously we denote 
\be
D=2n+\eps\,,
\ee
where $\varepsilon=0$  and
$\varepsilon=1$ for  even and odd number of dimensions, respectively. 
Operations $\flat$, $\sharp$ correspond to `lowering', `rising' of 
indices of vectors, forms, respectively. $\tens{\delta}$ denotes the
co-derivative. For a $p$-form  $\bs{\alpha}_p$ one has
$\tens{\delta\alpha}_p= \epsilon
\tens{*}\tens{d}\tens{*}\tens{\alpha}_p\,,$ where
$\tens{d}$ denotes the exterior derivative, $\tens{*}$ denotes the Hodge star
operator, and $\epsilon=(-1)^{p(D-p)+p-1}$. The `hook' operator $\hook$ denotes `contraction'. The scalar 
product of two vectors $\tens{a}$ and $\tens{b}$ is denoted by a dot; $\tens{a \cdot b}=\tens{a}\hook \tens{b}^\flat$.

{\bf Definition.} {\em A PCKY tensor $\tens{h}$  is a 
closed non-degenerate conformal Killing--Yano 2-form, ${\tens h}={1\over 2}\,h_{ab}\, \tens{d}x^a\!\wedge \tens{d}x^b.$ It obeys 
\be\label{PCKY}
\nabla_{X}\tens{h}=\tens{X}^{\flat}\wedge \tens{\xi}^\flat\, ,
\ee 
where $\tens{X}$ is an arbitrary vector field. 
}

The condition of {\em non-degeneracy} means that in a generic point of the manifold the skew symmetric matrix $h_{ab}$ 
has the (matrix) rank $2n$ and that the eigenvalues of $\tens{h}$ are 
functionally independent in some spacetime domain (see \cite{KrtousEtal:2008} for more details).
This means, that we exclude the possibility that 
$\tens{h}$ possesses constant eigenvalues, and in particular, that it is 
covariantly constant; $\tens{\xi}\neq 0$.
(For the discussion of cases when such degeneracies are admitted see \cite{HouriEtal:2008b, HouriEtal:2008c}.) 

The equation \eqref{PCKY} implies
\be\label{xi}
\tens{dh}=0\,,\quad \tens{\xi}^\flat=-\frac{1}{D-1}\tens{\delta h}\,.
\ee
It can be shown \cite{KrtousEtal:2008}, that for any spacetime admitting the PCKY tensor $\tens{h}$, $\tens{\xi}$ is a ({\em primary}) Killing vector. 
In tensor notations the definition \eq{PCKY} reads
\be\label{PCKY_compts}
\nabla_{c} h_{ab}=2g_{c[a}\xi_{b]},\quad
\xi_b=\frac{1}{D-1}\nabla_dh^{d}_{\ b}\,.
\ee

Let $\gamma$ be a null geodesic and $l^a=dx^a/d\tau$ be a tangent
vector to it; $\tau$ denotes the affine parameter. 
We denote the covariant derivative of
a tensor $\tens{T}$ along $\gamma$ by 
\begin{equation}
\tens{\dot T} =\nabla_l \tens{T}= l^a \nabla_a \tens{T}\,.
\end{equation}
In particular, $\tens{\dot l}=0$.

\section{Generic construction of a parallel-propagated frame}
In this section we outline the general construction of parallel-transported 
frames along {\em generic} geodesics.

\subsection{Construction of parallel-transported vectors $\tens{m}$ and $\tens{n}$}
Starting with $\tens{l}$ one can easily construct two additional 
parallel-propagated vectors. The construction is based on the following result:
Let $\tens{h}$ be a PCKY
tensor and $\tens{u}$ be a parallel-transported vector along a null geodesic $\tens{l}$, 
obeying $\tens{u \cdot l}=0$. Then the vector
\begin{equation}\label{w}
\tens{w}=(\tens{u}\hook \tens{h})^\sharp +\beta_{(u)} \tens{l}
\end{equation}
is parallel-transported along $\tens{l}$, provided that
\begin{equation}
\dot \beta_{(u)}=\tens{u \cdot \xi}\,.
\end{equation}
Here $\tens{\xi}$ is the primary Killing vector \eqref{xi}.

To prove this statement we use Eq. \eqref{PCKY} and the property of the hook operator. We find
\ba\label{proof}
\tens{\dot w}\!&=&\!(\tens{u}\hook \tens{\dot h})^\sharp +\dot \beta_{(u)} \tens{l}=
\bigr[\tens{u}\hook(\tens{l}^\flat\wedge \tens{\xi}^\flat)\bigr]^\sharp +\dot \beta_{(u)} \tens{l}\nonumber\\
\!&=&\!\tens{\xi}(\tens{u \cdot l})-
\tens{l}(\tens{u \cdot \xi})+\dot \beta_{(u)} \tens{l}=
\tens{l}\bigl[\dot \beta_{(u)}\!-\tens{u \cdot \xi}\bigr]\,.\nonumber
\ea

Obviously, $\tens{u}=\tens{l}$ obeys the requirements and
we may construct [$\beta = \beta_{(l)}$]
\begin{equation}\label{m}
\tens{m}= \frac{1}{\sqrt{-\kappa_1}}\bigl[(\tens{l}\hook \tens{h})^\sharp +\beta \tens{l}\bigr]\,.
\end{equation}
Here, $\kappa_1=Q_{ab}l^al^b$ is a constant of geodesic motion corresponding to the 
conformal Killing tensor $Q_{ab}=h_{ac}h_{\ b}^{c}$.
Since $\tens{\xi}$ is a primary Killing vector, $\dot \beta = 
\tens{l \cdot \xi}$ is also a constant and $\beta$ can be immediately integrated to get
\be\label{beta}
\beta=(\tens{l \cdot \xi})\tau\,.
\ee
We also find
\be\label{m_norm}
\tens{m \cdot l}=0\,,\quad
\tens{m \cdot m} = 1\,.
\ee
So, $\tens{m}$ given by \eqref{m} and \eqref{beta} is the normalized spacelike vector which is orthogonal to $\tens{l}$ and parallel-propagated along $\gamma$.

Taking $\tens{u}=\tens{m}$ in \eqref{w} we may construct another parallel-transported vector
\be\label{n}
\tens{n}= \tens{\tilde n}+\frac{1}{2}(\tens{{\tilde n}\cdot {\tilde n}})\,\tens{l}\,,
\ee
where
\begin{equation}\label{n_tilde}
\tens{\tilde n}=\frac{1}{\sqrt{-\kappa_1}}\bigl[(\tens{m}\hook \tens{h})^\sharp +\beta_{(m)} \tens{l}\bigr]\,,\ 
\dot \beta_{(m)}=\tens{m \cdot \xi}\,.
\end{equation}
We find
\be\label{N_norm}
\tens{n \cdot l}=-1\,,\quad
\tens{n \cdot m}=0\,,\quad
\tens{n \cdot n}=0\,.
\ee 
So, $\tens{n}$ is a null parallel-transported vector, external to $\tens{l}$ and orthogonal to $\tens{m}$. 
Moreover, we can easily show that it is independent of 
$\beta_{(m)}$. Indeed, using \eqref{n_tilde} we find that
\be\label{n_expr}
\begin{split}
\tens{n}&\,=\frac{1}{\sqrt{-\kappa_1}}(\tens{m}\hook \tens{h})^\sharp+
C\tens{l}\,,\\ 
C&\,= \frac{1}{2\kappa_1^2}\bigl(Q^{(2)}_{ab}l^al^b-\kappa_1\beta^2\bigr)\,,
\ Q^{(2)}_{ab}= Q_{ac}Q^c_{b}\,.
\end{split}
\ee
One might wonder whether it is not possible to generate more 
parallel-propagated vectors in this way. 
Unfortunately, the fact that $\tens{n}$ is external, $\tens{n \cdot l}=-1$, shows that $\tens{n}$ does not obey the requirements of the lemma anymore and we have to proceed differently.\footnote{In fact,
when $\tens{n}$ is used as a `seed' in \eqref{w} one obtains a new linear independent vector. This vector may be used as a seed again and so on---to produce 
the whole tower of linear independent vectors.
These vectors are, however, not parallel-transported. }

\subsection{Projection formalism: operator $F$} 
Let us consider the following 2-form $\tens{F}$:
\begin{equation}
F_{ab}=P_a^c P_b^d h_{cd}\,,
\end{equation}
where 
\be
P_{ab}=g_{ab}+2l_{(a}n_{b)}
\ee
is the projector to a $(D-2)$-dimensional space V, orthogonal to a 2-dimensional
space $U$ spanned by $\{\tens{l}, \tens{n}\}$. We have $P_{ab} l^b=0$, $P_{ab} n^b=0$.

Since vectors $\tens{l}$ and $\tens{n}$ are parallel-transported, so is 
$P_{ab}$; $\dot P_{ab}=0$. Therefore we find
\be
\dot F_{ab}=P_a^c P_b^d \dot h_{cd}=
2P_a^c P_b^d l_{[c}\xi_{d]}=0\,,
\ee 
where we have used Eq. \eqref{PCKY_compts}. So, the 2-form $\tens{F}$ is 
parallel-transported.
This implies that the eigenvalues of $\tens{F}$ as well as its Darboux subspaces 
are independently parallel-transported (see \cite{ConnellEtal:2008} for more details). 

The problem of finding remaining parallel-transported vectors along null geodesic $\gamma$ is now quite analogous to the problem of parallel transport along timelike (spacelike) geodesics. By solving the eigenvalue problem for the operator $\tens{F}^2$ one finds the eigenvectors spanning each of the Darboux subspaces of $\tens{F}$. In each Darboux subspace, the parallel-transported vectors are obtained by a $\tau$-dependent orthogonal transformation of these eigenvectors. 

The structure of the Darboux subspaces of $\tens{F}$ depends crucially on the chosen geodesic $\gamma$.
With increasing number of dimensions increases the number of degenerate cases
(corresponding to special geodesics) which require their own special treatment. 
In what follows we concentrate on generic geodesics for which the discussion significantly simplifies.

\subsection{Darboux subspace $V_0$}
We denote the Darboux subspace corresponding to the zeroth eigenvalue of $\tens{F}$ by $V_0$. For a generic geodesic $\gamma$, $V_0$ is 3-dimensional (4-dimensional) 
in the odd (even) number of spacetime dimensions. It is spanned by the parallel-transported vectors $\{\tens{l},\tens{n},\tens{m}\}$ and, in the even-dimensional case, $\tens{z}$ given by \eqref{z} below. 

The fact that $\tens{l}$ and $\tens{n}$ are zero-value eigenvectors of $\tens{F}$
is trivial. To prove that also $\tens{m}$ belongs to $V_0$ we first notice that, $\tens{m}$ being orthogonal to $\{\tens{l},\tens{n}\}$, is unaffected by projector $\tens{P}$, that is $P^b_a m^a=m^b$. Moreover, using \eqref{n_expr} we realize that $s^a=h^a_{\ b}m^b$ is a linear combination of $\tens{l}$ and $\tens{n}$, $\tens{s}=\alpha\tens{l}+\beta\tens{n}$, which when projected again by $\tens{P}$ gives zero. So we have 
\be
F_{ad}m^d=P_a^bh_{bc}P^c_d m^d=
P_a^bh_{bc}m^c=P_a^b s_b=0\,. 
\ee

In  an even number of spacetime dimensions we consider an additional vector 
$\tens{z}$ given by
\be\label{z}
\tens{z}=(\tens{l}\hook \tens{f})^\sharp\,,
\ee
where we have denoted  
\be
\tens{f}= \tens{*h}^{\wedge (n-1)}=\tens{*}(\!\!\!\!\!\underbrace{\tens{h}\wedge \ldots \wedge
\tens{h}}_{\mbox{\tiny{total of $(n-1)$ factors}}}\!\!\!\!\!)\, .
\ee
Tensor $\tens{f}$ is a Killing--Yano 2-form (see, e.g., \cite{KubiznakFrolov:2007}) and therefore vector $\tens{z}$ is automatically parallel-transported.
Forms $\tens{f}$ and $\tens{h}$ are related as 
\be
h^a{}_c f^c{}_b\propto\delta^a_b\,.
\ee
It then follows that $h^a{}_b z^b \propto l^a$. 
Using this relation one can easily show that $\tens{z}$ is 
spacelike parallel-transported vector, orthogonal to $\{\tens{l},\tens{n},\tens{m}\}$,
\be
\tens{z \cdot l}=0\,,\ \  \tens{z \cdot m}=0\,,\ \ 
\tens{z \cdot n}=0\,.
\ee
It can be normalized, so that, $\tens{z \cdot z}=1$.
Moreover, it is the last zero-value eigenvector of $\tens{F}$
spanning $V_0$. Indeed, we get
\be 
F^a{}_bz^b=P^a_dh^d{}_c P^c_b z^b=P^a_dh^d{}_cz^c\propto P^a_dl^d=0\,.\nonumber
\ee
To conclude, in the even-dimensional case, $V_0$ is spanned by explicitly constructed parallel-transported vectors  $\{\tens{l},\tens{n},\tens{m},\tens{z}\}$.

\subsection{Parallel-transported vectors in remaining Darboux subspaces} 
We restrict ourselves by considering generic geodesics for which all the remaining Darboux subspaces of $\tens{F}$ are two-dimensional.
Denote by $V_i$ the Darboux subspace of $\tens{F}$ corresponding to the
non-zero eigenvalue $-\lambda_i^2$ of the operator $\tens{F}^2$,
\be
\tens{F}^2\tens{v}=-\lambda_i^2 \tens{v}\,,\quad \tens{v}\in V_i\,,
\ee
and by $\{\bsi{}{n}_{\hat{i}} , \bsi{}{\tilde{n}}_{\hat{i}}\}$ the orthonormal basis in $V_i$.   This basis is related to the parallel-propagated orthonormal basis $\{\bsi{}{p}_{\hat{i}} , \bsi{}{\tilde{p}}_{\hat{i}}\}$ spanning $V_i$ by a 2D rotation
\be\label{pp}
\begin{split}
\bs{p}_{\hat{i}}=&\,\cos \gamma_{i} \bs{n}_{\hat{i}}-
\sin \gamma_{i} \bs{\tilde{n}}_{\hat{i}}\, \\
\bs{\tilde{p}}_{\hat{i}}=&\,\sin \gamma_{i} \bs{n}_{\hat{i}}+
\cos \gamma_{i} \bs{\tilde{n}}_{\hat{i}}\, ,\\
\dot{\gamma}_{i}=&\,\tens{\dot n}_{\hat{i}}\tens{\cdot}\tens{\,{\tilde{n}}}_{\hat{i}}=
-\tens{n}_{\hat{i}}\tens{\cdot \, \dot {\tilde{n}}}_{\hat{i}}\,.
\end{split}
\ee
If at the initial point $\tau=0$ bases $\{\bsi{}{p}_{\hat{i}} , \bsi{}{\tilde{p}}_{\hat{i}}\}$ and $\{\bsi{}{n}_{\hat{i}} , \bsi{}{\tilde{n}}_{\hat{i}}\}$ coincide,  
the initial conditions for Eqs. \eq{pp} are 
\begin{equation}\label{ic}
\gamma_{i}(\tau=0)=0\,.
\end{equation}

The construction of parallel-propagated vectors in a different Darboux subspace is 
exactly analogous, independent of the other constructions.

\section{Kerr-NUT-(A)dS spacetimes and their properties}

\subsection{Canonical metric element and Kerr-NUT-(A)dS spacetimes}

The most general canonical metric element admitting the PCKY tensor reads \cite{KrtousEtal:2008}
\begin{equation}\label{metrics}
\tens{g}=\!\sum_{\mu=1}^{n-1} (\tens{\omega}^{\hat \mu} 
\tens{\omega}^{\hat \mu}\! + \tens{\tilde \omega}^{\hat \mu} 
\tens{\tilde \omega}^{\hat \mu}) + \tens{\omega}^{\hat n} 
\tens{\omega}^{\hat n} -\tens{\tilde \omega}^{\hat n} 
\tens{\tilde \omega}^{\hat n}\!
+ \eps\, \tens{\omega}^{\hat \epsilon} \tens{\omega}^{\hat \epsilon}\,,
\end{equation}
where the basis 1-forms are ($\mu=1,\dots, n-1$)
\ba\label{one-forms}
\tens{\omega}^{\hat n} \!\!&=&\!\!\frac{\tens{d}r}{\sqrt{Q_n}}\,,\quad 
\tens{\omega}^{\hat \mu} = \frac{\tens{d}x_{\mu}}{\sqrt{Q_{\mu}}}\,,\nonumber\\
\tens{\tilde \omega}^{\hat n} \!\!&=&\!\! \sqrt{Q_{n}}
 \sum_{j=0}^{n-1}A_{n}^{(j)}\tens{d}\psi_j\;,\  
\tens{\tilde \omega}^{\hat \mu} = \sqrt{Q_{\mu}}
 \sum_{j=0}^{n-1}A_{\mu}^{(j)}\tens{d}\psi_j\;,\nonumber\\
\tens{\omega}^{\hat \epsilon} \!\!&=&\!\!\sqrt{Q_\epsilon}
\sum_{j=0}^nA^{(j)}\tens{d}\psi_j\;,\ \ Q_\epsilon=\frac{-c}{A^{(n)}}\,.
\ea
We enumerate the basis $\{\tens{\omega}\}$ so that 
$\tens{\tilde \omega}^{\hat n}$ is (the only one) timelike 1-form. 
Here, 
\ba\label{func}
A_{\mu}^{(j)}\!\!&=&\!\!\!\!\sum_{\ \substack{\nu_1<\dots<\nu_j\ \\
\nu_i\ne\mu}}\!\!\!\!\!x^2_{\nu_1}\dots x^2_{\nu_j},\quad\!\!\! 
A^{(j)}=\!\!\!\!\!\!\sum_{\nu_1<\dots<\nu_j}
\!\!\!\!\!x^2_{\nu_1}\dots x^2_{\nu_j}\;,\nonumber\\
Q_{\mu}\!\!&=&\!\frac{X_{\mu}}{U_{\mu}}\,,\ 
U_{\mu}=\prod_{\substack{\nu=1\\\nu\ne\mu}}^{n}(x_{\nu}^2-x_{\mu}^2)\,,\ 
x_n^2=-r^2\,,
\ea
and $X_\mu$, $X_n$ are arbitrary functions of $x_\mu$, $r$, respectively.
Time is denoted by $\psi_0$, azimuthal coordinates by $\psi_j$,
${j=1,\dots,m=D-n-1}$, $r$ is the Boyer-Lindquist type radial
coordinate, and ${x_\mu}$, ${\mu=1,\dots, n-1}$, stand for latitude
coordinates. 

The inverse metric reads
\be
\tens{g}^{-1}=\sum_{\mu=1}^{n-1} (\tens{e}_{\hat \mu} 
\tens{e}_{\hat \mu}\! + \tens{\tilde e}_{\hat \mu} 
\tens{\tilde e}_{\hat \mu}) + \tens{e}_{\hat n} 
\tens{e}_{\hat n} -\tens{\tilde e}_{\hat n} 
\tens{\tilde e}_{\hat n}\!
+ \eps\, \tens{e}_{\hat \epsilon} \tens{e}_{\hat \epsilon}\,,
\ee
where 
\be
\begin{split}
\tens{e}_{\hat n}=&\, \sqrt{Q_{n}}\pa_{r}\,,\quad 
\tens{e}_{\hat \mu}= \sqrt{Q_{\mu}}\pa_{x_{\mu}}\,,\\
\tens{\tilde e}_{\hat n}=&\,\frac{1}{\sqrt{X_nU_n}}
\sum_{j=0}^{m}r^{2(n-1-j)} \pa_{\psi_{j}}\,,\\
\tens{\tilde e}_{\hat \mu}=&\,\frac{1}{\sqrt{Q_{\mu}}U_{\mu}}
\sum_{j=0}^{m}(-x_\mu^2)^{n-1-j} \pa_{\psi_{j}}\,,\\
\tens{e}_{\hat \epsilon}=&\,-\frac{ \pa_{\psi_n}}{\sqrt{-cA^{(n)}}}\,.
\end{split}
\ee

The PCKY tensor for the canonical metric reads \cite{KubiznakFrolov:2007}, \cite{FrolovKubiznak:2007}
\begin{equation}\label{KYKNA}
\tens{h}=\sum_{\mu=1}^{n-1} x_\mu \tens{\omega}^{\hat \mu}
\wedge \tens{\tilde \omega}^{\hat \mu}
-r \tens{\omega}^{\hat n}\wedge \tens{\tilde \omega}^{\hat n}\,.
\end{equation}
Eqs. \eqref{metrics} and \eqref{KYKNA} mean that the basis $\{\tens{\omega}\}$ is an `orthogonal Darboux basis' of $\tens{h}$. We call such a basis a canonical one.
The canonical basis is fixed uniquely by the PCKY tensor up to 2D rotations
in each of the (KY) 2-planes 
$\tens{\omega}^{\hat n}\wedge \tens{\tilde \omega}^{\hat n}$\,,
$\tens{\omega}^{\hat \mu}\wedge \tens{\tilde \omega}^{\hat \mu}$.
The basis $\{\tens{\omega}\}$ is a special canonical basis for which 
many of the Ricci coefficients of rotation vanish \cite{HamamotoEtal:2007}, \cite{KrtousEtal:2008}.
We call it a {\em principal canonical} basis.
  
The PCKY tensor $\tens{h}$ generates the whole towers of explicit and hidden symmetries \cite{KrtousEtal:2007jhep}.
Namely, it generates all the isometries $\pa_{\psi_k}$, and in particular, 
the primary Killing vector $\tens{\xi}$, \eqref{xi}, 
\be\label{xi_metric}
\tens{\xi}=-\frac{1}{D-1}(\tens{\delta h})^\sharp=\pa_{\psi_0}\,.
\ee 
It also generates the set of the second-rank irreducible Killing tensors ($j=1,\dots,m$) 
\begin{equation}
\begin{split}\label{Kj}
\tens{K}^{(j)}=&\,\sum_{\mu=1}^{n-1} A_{\mu}^{(j)} 
(\tens{\omega}^{\hat \mu} \tens{\omega}^{\hat \mu}+
\tens{\tilde \omega}^{\hat \mu} \tens{\tilde \omega}^{\hat \mu})\\
+&\, A_{n}^{(j)} (\tens{\omega}^{\hat n} \tens{\omega}^{\hat n} 
-\tens{\tilde \omega}^{\hat n} \tens{\tilde \omega}^{\hat n})\!
+ \eps  A^{(j)}  \tens{\omega}^{\hat \epsilon} \tens{\omega}^{\hat \epsilon} \, .
\end{split}
\end{equation}
These objects are responsible for complete integrability of geodesic motion in the canonical spacetime.

When the vacuum Einstein equations with the cosmological constants are imposed,
\be\label{Rab}
{R_{ab}=(-1)^{n}(D-1)c_n\,g_{ab}}\,,
\ee
metric functions $X_\mu(x_\mu)$ and $X_n(r)$ take the following specific form \cite{HamamotoEtal:2007}:
\ba
X_{n}\!&=&\!-\sum\limits_{k=\varepsilon}^{n}c_k(-r^2)^k-2M r^{1-\varepsilon}
+\frac{\varepsilon c}{r^2}\,,\nonumber\\
X_{\mu}\!&=&\!\sum\limits_{k=\varepsilon}^{n}
c_kx_{\mu}^{2k}-2b_{\mu}x_{\mu}^{1-\varepsilon}
+\frac{\varepsilon c}{x_{\mu}^2}\,,
\ea
and the canonical element becomes the general Kerr-NUT-(A)dS
spacetime derived by Chen, L\"u, and Pope \cite{ChenEtal:2006cqg}.   
The parameter $c_n$ is proportional to the cosmological
constant  and the remaining constants $c_k$, $c>0$, and
$b_{\mu}$ are related to rotation parameters, mass, and NUT
parameters.

\subsection{Geodesics}
As we mentioned above, the geodesic motion in the canonical background \eqref{metrics}--\eqref{func} is  completely integrable \cite{PageEtal:2007, KrtousEtal:2007prd, HouriEtal:2008a}. In particular, 
the null geodesic velocity takes the following form \cite{KrtousEtal:2007jhep, FrolovKubiznak:2008}:
\begin{equation}
\tens{l}^\flat=\sum_{\mu=1}^n \bigl( l_{\hat \mu} \tens{\omega}^{\hat \mu} + {\tilde l}_{\hat \mu} \tens{\tilde \omega}^{\hat \mu}\bigr)
+\eps\,l_{\hat \epsilon} \tens{\omega}^{\hat \epsilon}\,,
\end{equation}
where 
\begin{equation}\label{basisvectors}
\begin{aligned}
l_{\hat n} =&\, \frac{\sigma_n}{(X_n U_n)^{1/2}}\,\bigl(
W_n^2 -X_n V_n\bigr)^{1/2}\,,\\
l_{\hat \mu} =&\, \frac{\sigma_\mu}{(X_\mu U_\mu)^{1/2}}\,\bigl(
X_\mu V_\mu-W_\mu^2 \bigr)^{1/2}\,,\\
{\tilde l}_{\hat n}=&\,\frac{W_n}{(X_n U_n)^{1/2}}\,,\quad 
{\tilde l}_{\hat\mu}=\frac{1}{\sqrt{Q_\mu}}\frac{W_\mu}{U_\mu}\,,\\
l_{\hat \epsilon} =&-\frac{\Psi_n}{\sqrt{-cA^{(n)}}}\,. 
\end{aligned}
\end{equation} 
Here, the constants $\sigma_\mu=\pm 1$ ($\mu=1,\dots, n$) are independent 
of one another and we have defined 
\begin{equation}\label{VW}
\begin{split}
V_n=&-\!\!\sum_{j=1}^{m}r^{2(n-1-j)}\kappa_j\,,\ 
V_\mu\!=\!\sum_{j=1}^{m}(-x_{\mu}^2)^{n-1-j}\kappa_j\,,\\  
W_n=&\!\!\sum_{j=0}^{m}r^{2(n-1-j)}\Psi_{\!j}\,,\ 
W_\mu\!=\!\sum_{j=0}^{m}(-x_{\mu}^2)^{n-1-j}\Psi_{\!j}\,.
\end{split}
\end{equation}
The quantities $\Psi_j$ and $\kappa_j$ are conserved and connected 
with the Killing vectors and the Killing tensors, respectively.
We also have 
\begin{equation}
\kappa_n=-\frac{\Psi_n^2}{c}\,.
\end{equation}
The coordinate components of the velocity are 
\be\label{xpsi_dot}
\begin{split}
\dot r=&\,\frac{\sigma_n}{U_n}\,\bigl(W_n^2-X_n V_n \bigr)^{1/2}\,,\\
\dot x_\mu=&\,\frac{\sigma_\mu}{|U_\mu|}\,\bigl(X_\mu V_\mu-W_\mu^2\bigr)^{1/2}\,,\\
\dot \psi_k=&\,\sum_{\mu=1}^{n-1}\frac{(-x_\mu^2)^{n-1-k}}{U_\mu X_\mu}\,W_\mu
-\frac{r^{2(n-1-k)}}{U_n X_n}\,W_n\\
&\,-\eps\frac{\Psi_n}{cA^{(n)}}\,\delta_{kn}\,.
\end{split}
\ee
One can symbolically integrate equations for $\psi_k\,$. 
Let $f$ be an arbitrary function of $r$ and $x_\nu$'s, obeying 
\begin{equation}\label{f}
\dot f=\frac{f_n(r)}{U_n}+
\sum_{\nu=1}^{n-1} \frac{f_\nu(x_\nu)}{U_\nu}\,.
\end{equation}
Then $f$ can be written as (see Appendix C in \cite{ConnellEtal:2008}) 
\begin{equation}\label{separ}
f=\!\int\!\!\frac{\sigma_n f_ndr}{\sqrt{W_n^2-X_n V_n}}
+\sum_{\nu=1}^{n-1}\int\!\!\frac{\sigma_\nu {\rm sign}(U_\nu) f_\nu dx_\nu}{\sqrt{X_\nu V_\nu-W_\nu^2}}\,.
\end{equation}
In particular, using the following identities (see, e.g., \cite{FrolovEtal:2007}):
\be
\begin{split}
\frac{1}{A^{(n)}}=&\,-\frac{1}{r^2U_n}+\sum_{\nu=1}^{n-1}\frac{1}{x_\nu^2U_\nu}\,,\\
1=&\,\frac{r^{2(n-1)}}{U_n}+\sum_{\mu=1}^{n-1}\frac{(-x_\mu^2)^{n-1}}{U_\mu}\,, 
\end{split}
\ee
we find that 
\ba\label{beta_m}
\psi_k\!\!&=&\!\!\!\int\!\!\!\frac{\sigma_n f_n^{(k)}dr}{\sqrt{W_n^2-X_n V_n}}+\!\sum_{\mu=1}^n\!\int\!\frac{\sigma_\mu{\rm sign}(U_\mu)f_\mu^{(k)}dx_\mu}{\sqrt{X_\mu V_\mu-W_\mu^2}}\,,\nonumber\\
f_n^{(k)}\!\!&=&-\frac{W_n}{X_n}r^{2(n-1-k)}+\eps\frac{\Psi_n}{c r^2}\,\delta_{kn}\,,
\nonumber\\
f_\mu^{(k)}\!\!&=&\!\!\frac{W_\mu}{X_\mu}(-x_\mu^2)^{n-1-k}-\eps\frac{\Psi_n}{c x_\mu^2}\,\delta_{kn}\,.
\ea
Similarly, we have
\be\label{tau}
\tau\!=\!\!\int\!\!\!\frac{\sigma_n r^{2(n-1)}dr}{\sqrt{W_n^2\!-\!X_n V_n}}+\!\!\sum_{\nu=1}^{n-1}\!\int\!\!\frac{\sigma_\nu {\rm sign}(U_\nu)(-\!x_\nu^2)^{n\!-\!1}dx_\nu}{\sqrt{X_\nu V_\nu\!-\!W_\nu^2}}\,.
\ee

\section{Parallel transport in Kerr-NUT-(A)dS spacetimes}

\subsection{Parallel-propagated frame}
We shall construct the parallel-propagated frame for a geodesic motion in four
steps. 
First, to simplify the calculations, we use the freedom of local 
2D rotations in the KY 2-planes of $\tens{h}$
to introduce the {\em velocity adapted} canonical basis in which 
$n$ components of the velocity vanish.
Next, we generate parallel-transported vectors $\tens{m}$, $\tens{n}$, and possibly $\tens{z}$.
In the third step, by studying the eigenvalue problem for the operator $\tens{F}^2$,
we find the orthonormal 1-forms $\{\tens{\varsigma}^{\hat i}, \tens{\tilde \varsigma}^{\,\hat i}\}$
spanning each of the 2D Darboux subspaces $V_i$.
Finally, in each $V_i$ we rotate these 1-forms by an (affine-parameter)-dependent rotation to obtain the (dual) parallel-transported frame.\footnote{%
In our setup it is somewhat more natural to work with 1-forms.
One could, of course, similarly construct the parallel-transported frame of vectors.} 

\subsubsection{Velocity  adapted canonical basis}
In order to construct the velocity adapted canonical basis we perform  the
boost transformation in the $\{\tens{\tilde \omega}^{\hat n},
\tens{\omega}^{\hat n}\}$ 2-plane and the rotation transformations in
each of the $\{\tens{\tilde \omega}^{\hat \mu}, \tens{\omega}^{\hat
\mu}\}$ 2-planes
\begin{equation}\label{newframe}
\begin{split}
\tens{\tilde o}^{\hat n}=&\,\cosh\alpha_n 
\tens{\tilde \omega}^{\hat n}+\sinh\alpha_n\tens{\omega}^{\hat n}\,,\\
\tens{o}^{\hat n}=&\,\sinh\alpha_n \tens{\tilde \omega}^{\hat n}
+\cosh\alpha_n\tens{\omega}^{\hat n}\,,\\
\tens{\tilde o}^{\hat \mu}=&\,\cos\alpha_\mu
\tens{\tilde \omega}^{\hat \mu}+\sin\alpha_\mu \tens{\omega}^{\hat \mu}\,,\\
\tens{o}^{\hat \mu}=&\,-\sin\alpha_\mu
\tens{\tilde \omega}^{\hat \mu}+\cos\alpha_\mu \tens{\omega}^{\hat \mu}\,,\\
\tens{o}^{\hat \epsilon}=&\ \tens{\omega}^{\hat \epsilon}\,.
\end{split}
\end{equation}
Here, we choose 
\begin{equation}\label{br}
\begin{split}
\cosh\alpha_n=&\,\frac{{\tilde l}_{\hat n}}{{\tilde k}_{\hat n}}\,,\quad
\sinh\alpha_n=\frac{l_{\hat n}}{{\tilde k}_{\hat n}}\,,\\
\cos\alpha_\mu=&\,\frac{{\tilde l}_{\hat \mu}}{{\tilde k}_{\hat \mu}}\,,\quad
\sin\alpha_\mu=\frac{l_{\hat \mu}}{{\tilde k}_{\hat \mu}}\,,
\end{split}
\end{equation}
and 
\begin{equation}\label{kn}
\begin{split}
{\tilde k}_{\hat n}=&\,-\sqrt{{\tilde l}_{\hat n}^2-l_{\hat n}^2}
=-\sqrt{\frac{V_n}{U_n}}\,,\\
{\tilde k}_{\hat \mu}=&\,\sqrt{{\tilde l}_{\hat\mu}^2+l_{\hat \mu}^2}
=\sqrt{\frac{V_\mu}{U_\mu}}\,.
\end{split}
\end{equation}
Such a transformation preserves the
form of the metric as well as the form of the PCKY tensor.
\ba\label{gh_velocityadapted}
\tens{g}\!&=&\!\sum_{\mu=1}^{n-1}(\tens{o}^{\hat \mu} \tens{o}^{\hat \mu}\! 
+\!\tens{\tilde o}^{\hat \mu} \tens{\tilde o}^{\hat \mu})\!
+\! \tens{o}^{\hat n} \tens{o}^{\hat n} \! -
\!\tens{\tilde o}^{\hat n} \tens{\tilde o}^{\hat n}\!+ 
\eps\, \tens{o}^{\hat \epsilon} \tens{o}^{\hat \epsilon},\nonumber\\
\tens{h}\!&=&\!\sum_{\mu=1}^{n-1} x_\mu 
\tens{o}^{\hat \mu}\wedge \tens{\tilde o}^{\hat \mu}
-r \tens{o}^{\hat n}\wedge \tens{\tilde o}^{\hat n}
\,.
\ea
Hence, the basis $\{\tens{o}\}$ is still canonical.
Moreover, one obtains the following form of the velocity:
\begin{equation}\label{l_velocityadapted}
\tens{l}^\flat=\sum_{\mu=1}^n {\tilde k}_{\hat \mu}\tens{\tilde o}^{\hat \mu}+
\eps l_{\hat \epsilon}\tens{o}^{\hat \epsilon}\,.
\end{equation} 
This form simplifies considerably 
the subsequent calculations, especially the task of solving the eigenvalue problem 
for $\tens{F}^2$.
We remark, that 
in the adapted basis $\{\tens{o}\}$ 
the components of the velocity 
depend on constants $\kappa_j$ only; 
the constants $\Psi_j$ and $\sigma_\mu$ are
absorbed in the definition of the new frame.

\subsubsection{Parallel-transported vectors in $V_0$}
The eigenspace $V_0$ is spread by $\tens{l}$, 
the vectors $\tens{m}$ and $\tens{n}$ given by \eqref{m} and
\eqref{n}, and, in an even number of spacetime dimensions,  by 
$\tens{z}$ \eqref{z}. 
Let us express these vectors in the velocity adapted basis \eqref{newframe}.
The vector $\tens{l}$ is given by \eqref{l_velocityadapted} and \eqref{kn}.
Using \eqref{l_velocityadapted} and \eqref{gh_velocityadapted}, we find
\ba
\tens{m}^\flat\!\!&=&\!\!\frac{1}{\sqrt{-\kappa_1}}\Bigl[
\sum\limits_{\mu=1}^{n-1}\bigl(
\tilde{k}_{\hat \mu}\beta \tens{\tilde o}^{\hat \mu}-\tilde{k}_{\hat \mu}x_\mu\tens{o}^{\hat \mu}\bigr)\nonumber \\
&\ +&\!\!\tilde{k}_{\hat n}\beta \tens{\tilde o}^{\hat n}-\tilde{k}_{\hat n}r\tens{o}^{\hat n} +\eps \beta l_{\hat \epsilon}\tens{o}^{\hat \epsilon}\Bigr]\,,\label{m_basis}\\
\tens{n}^\flat\!\!&=&\!\!
\sum\limits_{\mu=1}^{n-1}
\Bigl[\tilde{k}_{\hat \mu}\bigl(C+\frac{x_\mu^2}{\kappa_1}\bigr)\tens{\tilde o}^{\hat \mu}+\tilde{k}_{\hat \mu}\frac{\beta x_\mu}{\kappa_1}\tens{o}^{\hat \mu}\Bigr]\nonumber \\
&\ +&\!\!\tilde{k}_{\hat n}\bigl(C-\frac{r^2}{\kappa_1}\bigr)\tens{\tilde o}^{\hat n}+
\tilde{k}_{\hat n}\frac{\beta r}{\kappa_1}\tens{o}^{\hat n}
 +\eps C l_{\hat \epsilon}\tens{o}^{\hat \epsilon}\,,\ \ \label{n_basis} 
\ea
where
\be
C=\frac{1}{2\kappa_1^2}\Bigl(-r^4\tilde{k}_{\hat n}^{\,2}+\sum_{\mu=1}^{n-1} x_\mu^4 \tilde{k}_{\hat \mu}^{\,2}-\kappa_1\beta^2\Bigr)\,.
\ee
Moreover, using Eq. \eqref{xi_metric} we find  
\be
\dot\beta=\tens{l \cdot \xi}=
\tens{l \cdot} \pa_{\psi_0}=\Psi_0\,, \nonumber
\ee
and so 
\be\label{bm}
\beta=\Psi_0\tau\,.
\ee
Using Eq. \eqref{tau} one can express this angle as a function of $r$ and $x_\mu$'s. 

In an even number of spacetime dimensions we have an additional vector $\tens{z}$.
We find 
\be
\tens{f}\propto x_1\!\dots x_{n-1} \tens{o}^{\hat n}\wedge \tens{\tilde o}^{\hat n}
+\sum_{\mu=1}^{n-1}\! x_1\!\dots {\check x_\mu}\!\dots x_{n-1} 
\tens{o}^{\hat \mu}\wedge \tens{\tilde o}^{\hat \mu}.
\ee
Here, the symbol $\propto$ means equality up to a constant factor, and $\check x_\mu$ denotes that in the sum over $\mu$, $x_\mu$ is replaced by $r$. 
In consequence, we have the following expression for the (normalized) vector $\tens{z}$:
\be
\begin{split}
\tens{z}^\flat=\frac{1}{\sqrt{-\kappa_{n-1}}}&\,\Bigl(x_1\!\dots x_{n-1} {\tilde k}_{\hat n}\tens{o}^{\hat n}\\
&\,-
\sum_{\mu=1}^{n-1}\! x_1\!\dots {\check x_\mu}\!\dots x_{n-1} {\tilde k}_{\hat \mu} 
\tens{o}^{\hat \mu}\Bigr)\,. 
\end{split}
\ee

Using the transformation inverse to \eqref{newframe},
\begin{equation}\label{newframe_inverse}
\begin{split}
\tens{\tilde \omega}^{\hat n}=&\,\cosh\alpha_n 
\tens{\tilde o}^{\hat n}-\sinh\alpha_n\tens{o}^{\hat n}\,,\\
\tens{\omega}^{\hat n}=&\,-\sinh\alpha_n \tens{\tilde o}^{\hat n}
+\cosh\alpha_n\tens{o}^{\hat n}\,,\\
\tens{\tilde \omega}^{\hat \mu}=&\,\cos\alpha_\mu
\tens{\tilde o}^{\hat \mu}-\sin\alpha_\mu \tens{o}^{\hat \mu}\,,\\
\tens{\omega}^{\hat \mu}=&\,\sin\alpha_\mu
\tens{\tilde o}^{\hat \mu}+\cos\alpha_\mu \tens{o}^{\hat \mu}\,,\\
\tens{\omega}^{\hat \epsilon}=&\ \tens{o}^{\hat \epsilon}\,,
\end{split}
\end{equation}
one can easily obtain the above parallel-transported vectors in the principal 
basis $\{\tens{\omega}\}$.

\subsubsection{Darboux subspaces $V_i$}
Using \eqref{l_velocityadapted} and \eqref{n_basis} one can write down $\tens{F}$ in the velocity adapted basis. The general expression is quite involved and therefore we do not state it here. What is important is that  one can show that $\tens{F}$ is independent of $\beta$. Let us denote $\{\tens{\varsigma}\}$ the (dual) Darboux basis of $\tens{F}$.
Then, for a generic geodesic, we have
\be
\tens{F}=\sum_{i=1}^{n-2+\eps}\lambda_i \tens{\varsigma}^{\hat i}\wedge \tens{\tilde \varsigma}^{\,\hat i}\,.
\ee
Here, $\{\tens{\varsigma}^{\hat i}, \tens{\tilde \varsigma}^{\,\hat i}\}$ are orthonormal vectors spanning the Darboux subspace $V_i \,$, $\lambda_i>0$ are all different and correspond to the eigenvalues of $\tens{F}^2$;
$\tens{F}^2 \tens{v}_i=-\lambda_i^2\tens{v}_i, \ \tens{v}_i\in V_i$.
Obtaining the general form of $\{\tens{\varsigma}^{\hat i}, \tens{\tilde \varsigma}^{\,\hat i}\}$ is the biggest obstacle in writing down a general formula for the parallel-propagated basis in an arbitrary number of dimensions. Concrete examples are in the next section.

\subsubsection{Parallel-transported basis}
In order to construct parallel-transported vectors in each of the Darboux subspaces $V_i$ we perform the rotation
\be\label{pipi}
\begin{split}
\bs{\pi}^{\hat{i}}=&\,\cos \gamma_{i} \bs{\varsigma}^{\hat{i}}-
\sin \gamma_{i} \bs{\tilde{\varsigma}}^{\hat{i}}\, \\
\bs{\tilde{\pi}}^{\hat{i}}=&\,\sin \gamma_{i} \bs{\varsigma}^{\hat{i}}+
\cos \gamma_{i} \bs{\tilde{\varsigma}}^{\hat{i}}\, ,\\
\dot{\gamma}_{i}=&\,\tens{\dot \varsigma}^{\hat{i}}\tens{\cdot}\tens{\,{\tilde{\varsigma}}}^{\hat{i}}=
-\tens{\varsigma}^{\hat{i}}\tens{\cdot} \tens{\dot {\tilde{\varsigma}}}{}^{\,\hat{i}}\,,
\end{split}
\ee
with the initial conditions $\gamma_{i}(\tau=0)=0\,.$

When $\dot \gamma_i$ given by the last equation can be written in  the form 
\be\label{gamma_dot_sep}
\dot \gamma_i=M_i\Bigl[(r^2+\lambda^2_i)\prod_{\mu=1}^{n-1}(x_\mu^2-\lambda^2_i)\Bigr]^{-1}\!\!,
\ee
where $M_i$ is some constant, we can use the identity
\ba
\Bigl[(r^2\!\!&+&\!\!\lambda^2)\prod_{\mu=1}^{n-1}(x_\mu^2-\lambda^2)\Bigr]^{-1}\!\!\!=\nonumber\\
\!\!&&\!\!\frac{1}{(r^2+\lambda^2)U_n}-\sum_{\mu=1}^{n-1}\frac{1}{(x_\mu^2-\lambda^2)U_\mu}\,,
\ea
to symbolically integrate [cf. Eqs. \eqref{f}, \eqref{separ}]
\ba\label{beta_i}
\gamma_i\!\!&=&\!\!\!\int\!\!\!\frac{\sigma_n \gamma_n^{(i)} dr}{\sqrt{W_n^2-X_n V_n}}
-\!\sum_{\nu=1}^{n-1}\!\int\!\!\frac{\sigma_\nu {\rm sign}(U_\nu) \gamma_\nu^{(i)} dx_\nu}{\sqrt{X_\nu V_\nu-W_\nu^2}}\,,\nonumber\\
\gamma_n^{(i)}\!\!\!&=&\!\!\!\frac{M_i}{r^2+\lambda_i^2}\,,\quad 
\gamma_\nu^{(i)}=\frac{M_i}{x_\nu^2-\lambda_i^2}\,. 
\ea
We shall now give explicit examples of parallel-propagated frames
in $D=4, 5, 6$ canonical spacetimes.

\section{Special cases}

\subsection{Parallel transport in 4D}
The parallel transport along generic 
geodesics in the 4-dimensional canonical spacetime, derived earlier in \cite{Marck:1983null, KamranMarck:1986}, is from the point of view of the above described theory trivial. 
We write it only for completeness and because it encapsulates 
the important sub-case of parallel transport in the Carter's class of solutions \cite{Carter:1968cmp, Carter:1968pl}---describing among others a 4D rotating charged black hole in the cosmological background (see also Appendix C).

The metric reads
\begin{equation}\label{g4d}
\tens{g}=-\tens{\tilde \omega}^{\hat 2}\tens{\tilde \omega}^{\hat 2}+\tens{\omega}^{\hat 2}\tens{\omega}^{\hat 2}+
\tens{\tilde \omega}^{\hat 1}\tens{\tilde \omega}^{\hat 1}+\tens{\omega}^{\hat 1}\tens{\omega}^{\hat 1}\,,
\end{equation}
where  
\begin{equation}\label{tetrad}
\begin{split}
\tens{\tilde \omega}^{\hat 2}=&\,\sqrt{\frac{X_2}{U_2}}(\tens{d}\psi_0+x_1^2\tens{d}\psi_1 )\,,\ 
\tens{\omega}^{\hat 2}=\,\sqrt{\frac{U_2}{X_2}}\,\tens{d}r\,,\\ 
\tens{\tilde \omega}^{\hat 1}=&\,\sqrt{\frac{X_1}{U_1}}(\tens{d}\psi_0-r^2\tens{d}\psi_1 )\,,\ 
\tens{\omega}^{\hat 1}=\sqrt{\frac{U_1}{X_1}}\, \tens{d}x_1\,, 
\end{split}
\end{equation}
and $U_2=-U_1=x_1^2+r^2$.
The PCKY tensor is 
\be
\tens{h}=x_1\tens{\omega}^{\hat 1}\wedge\tens{\tilde \omega}^{\hat 1}-r\tens{\omega}^{\hat 2}\wedge\tens{\tilde \omega}^{\hat 2}\,.\label{KY4}
\ee
The components of the velocity are 
\be\label{u4}
\begin{split}
{\tilde l}_{\hat 2}=&\,\frac{W_2}{\sqrt{X_2U_2}}\,,\  
l_{\hat 2}=\frac{\sigma_2}{\sqrt{X_2U_2}}\sqrt{W_2^2-X_2V_2}\,,\\  
{\tilde l}_{\hat 1}=&\,\frac{-W_1}{\sqrt{X_1U_1}}\,,\  
l_{\hat 1}=\frac{\sigma_1}{\sqrt{X_1U_1}}\sqrt{X_1V_1-W_1^2}\,, 
\end{split}
\ee
where
\begin{equation}
\begin{split}
W_2=&\,\,r^2\Psi_0+\Psi_1\,,\quad
V_2=-\kappa_1>0\,,\\
W_1=&\,-x_1^2\Psi_0+\Psi_1\,,\quad
V_1=\kappa_1\,.
\end{split}
\end{equation}
In the velocity adapted frame $\{\tens{o}\}$, \eqref{newframe}, 
the parallel-transported frame reads 
\ba\label{vysl_4D}
\tens{l}^\flat\!\!\!&=&\!\!{\tilde k}_{\hat 1}(-\tens{\tilde o}^{\hat 2}+
\tens{\tilde o}^{\hat 1})\,,\ \ {\tilde k}_{\hat 1}=-{\tilde k}_{\hat 2}=\sqrt{\frac{V_2}{U_2}}=\sqrt{\frac{-\kappa_1}{U_2}}\,,\nonumber\\
\tens{m}^\flat\!\!\!&=&\!\!\frac{{\tilde k}_{\hat 1}}{\sqrt{-\kappa_1}}
\bigl(-\beta\tens{\tilde o}^{\hat 2}+r\tens{o}^{\hat 2}
+\beta\tens{\tilde o}^{\hat 1}-x_1\tens{o}^{\hat 1}\bigr)\,,\nonumber\\
\tens{n}^\flat\!\!\!&=&\!\!\frac{{\tilde k}_{\hat 1}}{\kappa_1}
\Bigl(\frac{U_2+\beta^2}{2}\tens{\tilde o}^{\hat 2}\!-\!\beta r\tens{o}^{\hat 2}
\!+\!\frac{U_2-\beta^2}{2}\tens{\tilde o}^{\hat 1}\!+\!\beta x_1\tens{o}^{\hat 1}\Bigr)\,,\nonumber\\
\tens{z}^\flat\!\!\!&=&\!\!\frac{{\tilde k}_{\hat 1}}{\sqrt{-\kappa_1}}(x_1\tens{o}^{\hat 2}+r\tens{o}^{\hat 1})\,,
\ea
where $\beta=\Psi_0\tau$, or, in terms of $r$ and $x_1$, 
\begin{equation}\label{beta4D}
\beta=\!\!\int\!\!\frac{\sigma_2 \Psi_0 r^2 dr}{\sqrt{W_2^2-X_2V_2}}+
\!\!\int\!\!\frac{\sigma_1 \Psi_0 x_1^2 dx_1}{\sqrt{X_1V_1-W_1^2}}\,.
\end{equation}

\subsection{Parallel transport in 5D}
Next, we consider the 5D canonical spacetime.
The metric reads
\begin{equation}\label{g5d}
\tens{g}=-\tens{\tilde \omega}^{\hat 2}\tens{\tilde \omega}^{\hat 2}+\tens{\omega}^{\hat 2}\tens{\omega}^{\hat 2}+
\tens{\tilde \omega}^{\hat 1}\tens{\tilde \omega}^{\hat 1}+\tens{\omega}^{\hat 1}
\tens{\omega}^{\hat 1}
+\tens{\omega}^{\hat \epsilon}\tens{\omega}^{\hat \epsilon}\,,
\end{equation}
where  
\begin{equation}\label{quintad}
\begin{split}
\tens{\tilde \omega}^{\hat 2}=&\,\sqrt{\frac{X_2}{U_2}}(\tens{d}\psi_0+x_1^2\tens{d}\psi_1 )\,,\  
\tens{\omega}^{\hat 2}=\,\sqrt{\frac{U_2}{X_2}}\,\tens{d}r\,,\\ 
\tens{\tilde \omega}^{\hat 1}=&\,\sqrt{\frac{X_1}{U_1}}(\tens{d}\psi_0-r^2\tens{d}\psi_1 )\,,\  
\tens{\omega}^{\hat 1}=\sqrt{\frac{U_1}{X_1}}\, \tens{d}x_1\,,\\
\tens{\omega}^{\hat \epsilon}=&\,\frac{\sqrt{c}}{rx_1}\,\left[\tens{d}\psi_0+
(x_1^2-r^2)\tens{d}\psi_1-x_1^2r^2\tens{d}\psi_2\right]\,,
\end{split}
\end{equation}
and $U_2=-U_1=x_1^2+r^2$.
The PCKY tensor is 
\be\label{KY5}
\tens{h}=x_1\tens{\omega}^{\hat 1}\wedge\tens{\tilde \omega}^{\hat 1}-r\tens{\omega}^{\hat 2}\wedge\tens{\tilde \omega}^{\hat 2}\,.
\ee
The components of the velocity are 
\ba\label{u5}
{\tilde l}_{\hat 2}\!&=&\!\frac{W_2}{\sqrt{X_2U_2}}\,,\  
l_{\hat 2}=\frac{\sigma_2}{\sqrt{X_2U_2}}\sqrt{W_2^2-X_2V_2}\,,\nonumber\\  
{\tilde l}_{\hat 1}\!&=&\!\frac{-W_1}{\sqrt{X_1U_1}}\,,\  
l_{\hat 1}=\frac{\sigma_1}{\sqrt{X_1U_1}}\sqrt{X_1V_1-W_1^2}\,,\nonumber\\
l_{\hat \epsilon}\!&=&\!-\frac{\Psi_2}{\sqrt{c}x_1r}\,, 
\ea
where
\begin{equation}\label{W5}
\begin{split}
W_1=&\,-x_1^2\Psi_0+\Psi_1-\frac{\Psi_2}{x_1^2}\,,\ 
V_1=\kappa_1+\frac{\Psi_2^2}{cx_1^2}\,,\\
W_2=&\,r^2\Psi_0+\Psi_1+\frac{\Psi_2}{r^2}\,,\ 
V_2=-\kappa_1+\frac{\Psi_2^2}{cr^2}\,.
\end{split}
\end{equation}
In the velocity adapted frame $\{\tens{o}\}\,$,  \eqref{newframe},
we have 
\ba
\tens{l}^\flat\!&=&\!{\tilde k}_{\hat 2}\tens{\tilde o}^{\hat 2}\!+{\tilde k}_{\hat 1}\tens{\tilde o}^{\hat 1}\!+l_{\hat \epsilon}\tens{o}^{\hat \epsilon}\,,\nonumber\\ 
\tens{m}^\flat\!\!\!&=&\!\!\!\!\frac{1}{\sqrt{\!-\!\kappa_1}}
\bigl(\beta{\tilde k}_{\hat 2}\tens{\tilde o}^{\hat 2}\!\!-\!r{\tilde k}_{\hat 2}\tens{o}^{\hat 2}\!
\!+\!\beta {\tilde k}_{\hat 1}\tens{\tilde o}^{\hat 1}\!\!-\!x_1{\tilde k}_{\hat 1}\tens{o}^{\hat 1}\!\!+\!\beta l_{\hat \epsilon} \tens{o}^{\hat \epsilon}\bigr),\nonumber\\
\tens{n}^\flat\!\!\!&=&{\tilde k}_{\hat 2}\Bigl(C-\frac{r^2}{\kappa_1}\Bigr)\tens{\tilde o}^{\hat 2}\!+\!{\tilde k}_{\hat 2}\frac{\beta r}{\kappa_1}\,\tens{o}^{\hat 2}
\!+\! {\tilde k}_{\hat 1}\Bigl(C+\frac{x_1^2}{\kappa_1}\Bigr)\tens{\tilde o}^{\hat 1}\nonumber\\
\!\!\!&\quad &\ +{\tilde k}_{\hat 1}\frac{\beta x_1}{\kappa_1}\,\tens{o}^{\hat 1}\!+\!Cl_{\hat \epsilon} \tens{o}^{\hat \epsilon}\,, 
\ea
where
\ba\label{u5c}
{\tilde k}_{\hat 2}\!&=&\!-\sqrt{\frac{V_2}{U_2}}\,,\quad 
{\tilde k}_{\hat 1}=\sqrt{\frac{V_1}{U_1}}\,,\nonumber\\
C\!&=&\!\frac{1}{2\kappa_1^2}\bigl(-r^4{\tilde k}_{\hat 2}^2+x_1^4{\tilde k}_{\hat 1}^2-\kappa_1\beta^2\bigr)\,,\\
\beta\!&=&\!\!\!\int\!\!\frac{\sigma_2 \Psi_0 r^2 dr}{\sqrt{W_2^2-X_2V_2}}+
\!\!\int\!\!\frac{\sigma_1 \Psi_0 x_1^2 dx_1}{\sqrt{X_1V_1-W_1^2}}\,\,.\nonumber
\ea
The 2-form $\tens{F}$ reads
\begin{equation}
\tens{F}=\lambda\, \tens{\varsigma} \wedge \tens{\tilde \varsigma}\,,\quad
\lambda=\frac{|\Psi_2|}{\sqrt{-c\kappa_1}}\,
\end{equation}
where
\ba
\tens{\varsigma}\!\!&=&\!\!\frac{\sqrt{(r^2\!+\!\lambda^2)(x_1^2\!-\!\lambda^2)}}{rx_1}\Bigl(\!-\frac{x_1 F_r}{\sqrt{U_2}}\,\tens{\tilde o}^{\hat 2}\!\!+\!\frac{rF_{x_1}}{\sqrt{U_2}}\,\tens{\tilde o}^{\hat 1}\!\!+\!\tens{o}^{\hat \epsilon}\!\Bigr),\nonumber\\
\tens{\tilde \varsigma}\!\!&=&\!\!\frac{\sqrt{r^2+\lambda^2}}{\sqrt{U_2}}\Bigl(\!-\frac{\lambda^2}{r^2+\lambda^2}\frac{1}{F_rF_{x_1}}\,\tens{o}^{\hat 2}\!+\!\tens{o}^{\hat 1}\Bigr)\,.
\ea 
Here we have introduced
\begin{equation}
F_r=\frac{x_1 l_{\hat \epsilon}}{\sqrt{V_2}}\,,\quad 
F_{x_1}=-\frac{rl_{\hat \epsilon}}{\sqrt{-V_1}}\,, 
\end{equation}
which are functions of $r$, $x_1$, respectively.
Using \eqref{pipi} we find
\be
\dot \gamma=\!\frac{M}{(x_1^2\!-\!\lambda^2)(r^2\!+\!\lambda^2)}\,,\ 
M\!=\!\frac{\lambda^2\Psi_2\Psi_0\!-\!\Psi_2\Psi_1\!-\!c\kappa_1}{\sqrt{-c\kappa_1}}\,,
\ee
which is of the form \eqref{gamma_dot_sep}. Therefore, the 
parallel-propagated forms $\{\tens{\pi}, \tens{\tilde \pi}\}$ are given by \eqref{pipi},
where 
\ba\label{beta1}
\gamma \!&=&\!\int\!\!\frac{\sigma_2 \gamma_r dr}{\sqrt{W_2^2-X_2 V_2}}
+\!\int\!\!\frac{\sigma_{1} \gamma_{x_1} dx_1}{\sqrt{X_1 V_1-W_1^2}}\,,\nonumber\\
\gamma_r\!&=&\!\frac{M}{r^2+\lambda^2}\,,\ \ 
\gamma_{x_1}=\frac{M}{x_1^2-\lambda^2}\,. 
\ea

\subsection{Parallel transport in 6D}
Finally we consider the 6D canonical spacetime.
The metric reads
\ba\label{g6d}
\tens{g}\!\!&=&\!\!-\tens{\tilde \omega}^{\hat 3}\tens{\tilde \omega}^{\hat 3}+\tens{\omega}^{\hat 3}\tens{\omega}^{\hat 3}+\sum_{\mu=1}^2\bigr(
\tens{\omega}^{\hat \mu}\tens{\omega}^{\hat \mu}+
\tens{\tilde \omega}^{\hat \mu}\tens{\tilde \omega}^{\hat \mu}\bigl)\,,\nonumber\\
\tens{\omega}^{\hat 3}\!\!\!&=&\!\!\sqrt{\frac{U_3}{X_3}}\,\tens{d}r\,,\, 
\tens{\omega}^{\hat 2}\!=\!\sqrt{\frac{U_2}{X_2}}\,\tens{d}x_2\,,\, 
\tens{\omega}^{\hat 1}\!=\!\sqrt{\frac{U_1}{X_1}}\, \tens{d}x_1\,, \nonumber\\
\tens{\tilde \omega}^{\hat 3}\!\!\!&=&\!\!\sqrt{\frac{X_3}{U_3}}\left(\tens{d}\psi_0+
A_r\tens{d}\psi_1+x_1^2x_2^2\tens{d}\psi_2\right)\,,\nonumber\\
\tens{\tilde \omega}^{\hat 2}\!\!\!&=&\!\!\sqrt{\frac{X_2}{U_2}}\left(\tens{d}\psi_0+
A_{x_2}\tens{d}\psi_1-x_1^2r^2\tens{d}\psi_2\right)\,,\nonumber\\
\tens{\tilde \omega}^{\hat 1}\!\!\!&=&\!\!\sqrt{\frac{X_1}{U_1}}\left(\tens{d}\psi_0+
A_{x_1}\tens{d}\psi_1-x_2^2r^2\tens{d}\psi_2\right)\,.
\ea
Here  
\ba
A_r\!\!&=&\!x_1^2+x_2^2\,,\ 
A_{x_1}\!=x_2^2-r^2\,,\ A_{x_2}\!=x_1^2-r^2\,,\nonumber\\
U_3\!\!&=&\!(x_1^2+r^2)(x_2^2+r^2)\,,\nonumber\\
U_2\!\!&=&\!-(x_2^2+r^2)(x_1^2-x_2^2)\,,\nonumber\\
U_1\!\!&=&\!(x_1^2+r^2)(x_1^2-x_2^2)\,.\nonumber
\ea
The PCKY tensor is 
\be\label{KY6}
\tens{h}=x_1\tens{\omega}^{\hat 1}\wedge\tens{\tilde \omega}^{\hat 1}+
x_2\tens{\omega}^{\hat 2}\wedge\tens{\tilde \omega}^{\hat 2}-r\tens{\omega}^{\hat 3}\wedge\tens{\tilde \omega}^{\hat 3}\,.
\ee
The components of the velocity are 
\ba\label{u6}
{\tilde l}_{\hat 3}\!\!\!&=&\!\!\frac{W_3}{\sqrt{X_3U_3}}\,,\  
l_{\hat 3}\!=\!\frac{\sigma_3}{\sqrt{X_3U_3}}\sqrt{W_3^2-X_3V_3}\,,\nonumber\\  
{\tilde l}_{\hat 2}\!\!\!&=&\!\!-\frac{W_2}{\sqrt{X_2U_2}}\,,\  
l_{\hat 2}\!=\!\frac{\sigma_2}{\sqrt{X_2U_2}}\sqrt{X_2V_2-W_2^2}\,,\nonumber\\  
{\tilde l}_{\hat 1}\!\!\!&=&\!\!\frac{W_1}{\sqrt{X_1U_1}}\,,\  
l_{\hat 1}\!=\!\frac{\sigma_1}{\sqrt{X_1U_1}}\sqrt{X_1V_1-W_1^2}\,,
\ea
where
\begin{equation}\label{W6}
\begin{split}
W_3=&\,r^4\Psi_0+r^2\Psi_1+\Psi_2\,,\ 
V_3=-r^2\kappa_1-\kappa_2\,,\\
W_2=&\,x_2^4\Psi_0-x_2^2\Psi_1+\Psi_2\,,\ 
V_2=-x_2^2\kappa_1+\kappa_2\,,\\
W_1=&\,x_1^4\Psi_0-x_1^2\Psi_1+\Psi_2\,,\ 
V_1=-x_1^2\kappa_1+\kappa_2\,.
\end{split}
\end{equation}
The parallel-transported vectors spanning $V_0$ are 
\ba
\tens{l}^\flat\!&=&\!{\tilde k}_{\hat 3}\tens{\tilde o}^{\hat 3}+{\tilde k}_{\hat 2}\tens{\tilde o}^{\hat 2}+{\tilde k}_{\hat 1}\tens{\tilde o}^{\hat 1}\,,\nonumber\\ 
\tens{m}^\flat\!\!\!&=&\!\!\frac{1}{\sqrt{-\kappa_1}}
\Bigl(\beta {\tilde k}_{\hat 3}\tens{\tilde o}^{\hat 3}-r{\tilde k}_{\hat 3}\tens{o}^{\hat 3}+\beta{\tilde k}_{\hat 2}\tens{\tilde o}^{\hat 2}\nonumber\\
&&\quad\quad -x_2{\tilde k}_{\hat 2}\tens{o}^{\hat 2}+\beta {\tilde k}_{\hat 1}\tens{\tilde o}^{\hat 1}-x_1 {\tilde k}_{\hat 1}\tens{o}^{\hat 1} \Bigr)\,,\nonumber\\
\tens{n}^\flat\!\!&=&\!\!
\tilde{k}_{\hat 3}\bigl(C-\frac{r^2}{\kappa_1}\bigr)\tens{\tilde o}^{\hat 3}+
\tilde{k}_{\hat 3}\frac{\beta r}{\kappa_1}\,\tens{o}^{\hat 3}\\
&&+
\sum\limits_{\mu=1}^{2}
\Bigl[\tilde{k}_{\hat \mu}\bigl(C+\frac{x_\mu^2}{\kappa_1}\bigr)\tens{\tilde o}^{\hat \mu}+\tilde{k}_{\hat \mu}\frac{\beta x_\mu}{\kappa_1}\,\tens{o}^{\hat \mu}\Bigr]\,,\nonumber \\
\tens{z}^\flat\!\!&=&\!\!\frac{1}{\sqrt{-\kappa_2}}
\Bigl(x_1x_2{\tilde k}_{\hat 3}\tens{o}^{\hat 3}\!-\!rx_1{\tilde k}_{\hat 2}\tens{o}^{\hat 2}\!-\!rx_2{\tilde k}_{\hat 1}\tens{o}^{\hat 1}\Bigr)\,,\nonumber
\ea
where
\ba\label{u6c}
{\tilde k}_{\hat 3}\!&=&\!-\sqrt{\frac{V_3}{U_3}}\,,\ 
{\tilde k}_{\hat 2}=\sqrt{\frac{V_2}{U_2}}\,,\  
{\tilde k}_{\hat 1}=\sqrt{\frac{V_1}{U_1}}\,,\nonumber\\
C\!&=&\!\frac{1}{2\kappa_1^2}\bigl(-r^4\tilde{k}_{\hat n}^{\,2}+x_2^4 \tilde{k}_{\hat 2}^{\,2}+x_1^4 \tilde{k}_{\hat 1}^{\,2}-\kappa_1\beta^2\bigr)\,,\nonumber\\
\beta\!&=&\!\!\int\!\!\frac{\Psi_0\sigma_3 r^4 dr}{\sqrt{W_3^2-X_3V_3}}-
\int\!\!\frac{\Psi_0 \sigma_2  x_2^4 dx_2}{\sqrt{X_2 V_2-W_2^2}}\nonumber\\
&&+\int\!\!\frac{\Psi_0 \sigma_1  x_1^4 dx_1}{\sqrt{X_1 V_1-W_1^2}}\,.\nonumber
\ea
The 2-form $\tens{F}$ reads
\begin{equation}
\tens{F}=\lambda\, \tens{\varsigma} \wedge \tens{\tilde \varsigma}\,,\quad
\lambda=\sqrt{\frac{\kappa_2}{\kappa_1}}\,,
\end{equation}
where
\ba
\tens{\varsigma}\!\!&=&\!\!\sqrt{\frac{(r^2\!+\!\lambda^2)(\lambda^2\!-\!x_2^2)}{U_1}}
\,\bigl(F_1\tens{\tilde o}^{\hat 3}\!+\!F_2\tens{\tilde o}^{\hat 2}\!+\!\tens{\tilde o}^{\hat 1}\bigr)\,,\\
\tens{\tilde \varsigma}\!\!&=&\!\!\frac{1}{\lambda}\sqrt{\frac{(r^2\!+\!\lambda^2)(\lambda^2\!-\!x_2^2)}{U_1}}
\bigl(rF_1\tens{o}^{\hat 3}\!+\!x_2F_2\tens{o}^{\hat 2}\!+\!x_1\tens{o}^{\hat 1}\bigr)\,.\nonumber
\ea
Here we have introduced
\begin{equation}
F_1=\frac{U_1{\tilde k}_{\hat 1}{\tilde k}_{\hat 3}}{\kappa_1 (r^2+\lambda^2)}\,,\quad
F_2=\frac{U_1{\tilde k}_{\hat 2}{\tilde k}_{\hat 1}}{\kappa_1(\lambda^2-x_2^2)}\,.
\end{equation}
Using Eq. \eqref{pipi} we find
\be
\dot \gamma =\frac{-\lambda\,(\Psi_2-\lambda^2\Psi_1+\lambda^4\Psi_0)}{(x_1^2-\lambda^2)(x_2^2-\lambda^2)(r^2+\lambda^2)}\,.
\ee
This means that also in 6D the angle $\gamma$ can be symbolically integrated---it is given by \eqref{beta_i}---and the parallel-transported forms $\{\tens{\pi}, \tens{\tilde \pi}\}$, \eqref{pipi}, explicitly constructed.

\section{Principal null directions}
So far we have described the construction of a parallel-propagated frame along generic null geodesics. Similar to the timelike case, with increasing number of spacetime dimensions increases the number of degenerate cases for which this construction has to be modified. One type of degeneracy occurs for special geodesics for which the spectrum of the operator $\tens{F}$ is degenerate. These geodesics are characterized by a special choice of constants motion and the parallel transport along them was  partly described in \cite{ConnellEtal:2008}. 
In this section, we concentrate on 
a more fundamental degeneracy which happens when the very construction of the external vector $\tens{n}$, \eqref{n}, fails. 
Such a degeneracy occurs for the velocity vector $\tens{l}$ which is the eigenvector of the PCKY tensor. 
In this case one cannot proceed with constructing the 2-form $\tens{F}$ and the whole procedure described in Section II breaks down.
In the following subsection we show that such a situation occurs for an important class of geodesics called the {\em principal null directions}. 
The parallel-propagated frame along these directions is described in the next subsection.

\subsection{Principal null directions as the eigenvectors of the PCKY tensor} 
The {\em principal null directions} (the Weyl aligned null geodesics, see, e.g., \cite{Coley:2008}) play an important role in many physical situations. 
In a spacetime admitting the PCKY tensor the principal null directions, $\tens{l}_{\pm}$, coincide with the (real) eigenvectors of the PCKY tensor,
\be\label{ee}
\tens{l}_{\pm}\hook\tens{h}=\pm \lambda \tens{l}^\flat_{\pm}\,,
\ee
see \cite{MasonTaghavi:2008} or Appendix C.1 in \cite{Kubiznak:phd}.

Namely, for the canonical metric element the affine-parametrized principal null directions are \cite{HamamotoEtal:2007, PravdaEtal:2007}
\begin{equation}\label{momcoor}
\begin{split}
  \tens{l}_{\pm} &= \frac1{\sqrt{Q_n}}\,\bigl(\tens{\tilde e}_{\hat n}\pm\tens{e}_{\hat n}\bigr)\\
       &=\pm\cv{r} +\frac1{X_n}\sum_{j=0}^{m} r^{2(n{-}1{-}j)}\cv{\psi_j}\;.
\end{split}
\end{equation}
These directions are characterized by the following constants of motion: 
\begin{equation}\label{intconst}
  \kappa_j=0\;,\quad\Psi_j=-A^{(j)}_n\;,\quad\Psi_n=0\;,
\end{equation}
where ${j=0,\dots,n-1}$ and the constant ${\Psi_n}$ is relevant only in odd dimensions.
Eq.~\eqref{momcoor} gives also coordinate components of $\tens{l}_{\pm}$, which lead
to the equations for geodesics
\begin{equation}\label{geodeqcoor}
  \dot r = \pm1\;,\quad \dot x_\mu = 0\;,\quad \dot\psi_j = \frac{r^{2(n{-}1{-}j)}}{X_n}\;,
\end{equation}
where ${\mu=1,\dots,n-1}$ and ${j=0,\dots,m}$. These can be integrated to get \begin{equation}\label{geod}
  r=\pm\tau\;,\quad \psi_j=\pm\int_0^{\pm\tau}\frac{r^{2(n{-}1{-}j)}}{X_n}\,dr + \psi_j^{(0)}\;.
\end{equation}

\subsection{Parallel transport}
Now, we turn to the task of parallel transport along the principal null direction $\tens{l}_{+}$.\footnote{The parallel transport along $\tens{l}_-$ is analogous.} 
The parallel-propagated frame can be obtained by a 
sequence of local Lorentz transformations of the canonical basis 
$\{\tens{l}, \tens{n}, \tens{e}_{\hat \mu}, \tens{\tilde e}_{\hat \mu}, 
\tens{e}_{\hat \epsilon} \}$, 
\be
\tens{l}=\frac{1}{\sqrt{2}}\bigl(\tens{\tilde e}_{\hat n}+\tens{e}_{\hat n}\bigr)\,,\quad \tens{n}=\frac1{\sqrt{2}}\bigl(\tens{\tilde e}_{\hat n}-\tens{e}_{\hat n}\bigr) \,,
\ee
along the geodesic (see Appendix B). Here, $\mu=1,\dots,n-1$, and $\tens{e}_{\hat \epsilon}$ is relevant only in an odd number of spacetime dimensions. The resulting parallel-propagated frame is
$\{\pfr\tens{l}, \pfr\tens{n}, \pfr\tens{e}_{\hat \mu},\pfr\tens{\tilde e}_{\hat \mu}, \pfr\tens{e}_{\hat \epsilon } \}$\,, where
\begin{equation}\label{partr-KY}
\begin{aligned}
 \pfr\tens{l}&=\frac1{\sqrt{\smash[b]{Q_n}}}\,\lv\;,\\
  \pfr\tens{n}&=\sqrt{\smash[b]{Q_n}}\,\tens{n}+\sum_{\mu=1}^{n-1}\sqrt{2}\sqrt{\smash[b]{Q_\mu}}\tens{\tilde e}_{\hat \mu}+\eps\sqrt{2}\sqrt{\smash[b]{Q_\epsilon}}\tens{e}_{\hat \epsilon}\\
&\quad+\Bigl(\sum_{\mu=1}^{n-1}Q_\mu+\eps\, Q_\epsilon\Bigr)\frac1{\sqrt{\smash[b]{Q_n}}}\,\tens{l}\;,\\
\pfr\tens{e}_{\hat \mu}&=\frac{x_\mu}{\sqrt{\smash[b]{x_\mu^2{+}r^2}}}\,\tens{e}_{\hat \mu}-\frac{r}{\sqrt{\smash[b]{x_\mu^2{+}r^2}}}\,
                 \biggl(\tens{\tilde e}_{\hat \mu}{+}\sqrt{2}\frac{\sqrt{\smash[b]{Q_\mu}}}{\sqrt{\smash[b]{Q_n}}}\,\lv\biggr)\;,\\[0.5ex]
  \pfr\tens{\tilde e}_{\hat \mu}&=\frac{r}{\sqrt{\smash[b]{x_\mu^2{+}r^2}}}\,\tens{e}_{\hat \mu}+\frac{x_\mu}{\sqrt{\smash[b]{x_\mu^2{+}r^2}}}\,
                 \biggl(\tens{\tilde e}_{\hat \mu}{+}\sqrt{2}\frac{\sqrt{\smash[b]{Q_\mu}}}{\sqrt{\smash[b]{Q_n}}}\,\lv\biggr)\;,\\[0.5ex]
  \pfr\tens{e}_{\hat \epsilon}&=\tens{e}_{\hat \epsilon}+\sqrt{2}\,\frac{\sqrt{\smash[b]{Q_\epsilon}}}{\sqrt{\smash[b]{Q_n}}}\,\lv\;.
\end{aligned}
\end{equation}
Similar to 4D, it may be convenient to find a parallel-propagated complex null frame.
This is done in Appendix B.

\section{Conclusions} 
 
In this paper we have studied the equations describing the parallel-transport along null geodesics in spacetimes which possess a (non-degenerate) principal conformal
Killing-Yano (PCKY) tensor. 
When the vacuum Einstein equations with the cosmological constant are imposed,  this class of metrics coincides with the Kerr-NUT-(A)dS
spacetimes, describing the higher-dimensional rotating black holes with
NUT parameters, in an asymptotically flat or (A)dS background. In
particular, a solution of this problem gives effective tools for studying
the polarization of light beams in backgrounds of considered black hole metrics. 

A tangent vector to the null ray, which is evidently parallel-propagated,
determines a $(D-1)$-dimensional null plane to
which it is orthogonal.  Our main observation is, that using the PCKY tensor
one can obtain a parallel-propagated along the null geodesic
vector which does not belong to this null plane. We used these two
parallel-propagated vectors to construct a projection operator on
a $(D-2)$-dimensional subspace. By using the eigenvectors of the PCKY tensor
projected to this subspace we found two-dimensional Darboux planes invariant
under the parallel-transport, and by proper rotations in these planes
we constructed the required parallel-propagated basis. Though the idea
of this construction is rather simple, concrete calculations in 
higher-dimensional spacetimes are quite involved. We performed them concretely 
in spacetimes with $D\le 6$. In each of these cases the
final first order ordinary differential equations specifying rotations
in the 2D Darboux planes were solved by the separation of variables.
We expect that this remains true for any number of dimensions. 

The class of the Kerr-NUT-(A)dS spacetimes we considered in this
paper belongs to the algebraic type D. These spacetimes possess
two special congruences of null geodesics called the principal null
directions. Tangent vectors to these geodesics are `eigenvectors' of
the Weyl tensor. At the same time they are null
eigenvectors of the PCKY tensor. This property implies that for this
subclass of null geodesics the problem of the parallel-transport
becomes degenerate and requires special consideration. We have studied
this degenerate case and directly solved the corresponding equations of
parallel transport. This result might be useful for studying the
peeling-off property in the Kerr-NUT-(A)dS spacetimes.
   
\appendix

\section{Geometric optics in higher-dimensional spacetimes}

It is well known that if the wave length of massless field radiation
is much smaller than a characteristic scale on which the
gravitational field changes one can use the geometric optics
approximation. In this approximation a normal to a surface of
constant phase is a null vector which is tangent to a null geodesic
describing a motion of a massless quantum. We collect here useful
relations of the geometric optics in a higher dimensional curved
spacetime. To make the presentation concrete we discuss the
electromagnetic field propagation. We closely follow the nice presentation of
the MTW book \cite{MTW}, which requires only tiny changes connected
with the number of dimensions $D$ which is now not four but arbitrary.
Maxwell equations in a spacetime with the metric $g_{ab}$,
($a,b=0,\ldots,D-1$) in the Lorentz gauge have the form
\ba
&&\nabla^b\nabla_b A^{a}-R^a_b A^b=0\, ,\n{eqn}\\
&&\nabla_a A^{a}=0\, .\n{lor}
\ea
We write the potential $A_{a}$ in the form
\be\n{gop}
A_a=\Re \Bigl\{\bigl[{\cal A}+O(\epsilon)\bigr]_a\, e^{iS/\epsilon}\Bigr\}\, .
\ee
Here $\epsilon$ is a small parameter.

Substituting \eq{gop} into \eq{lor} and keeping the term of the
leading order $\epsilon^{-1}$ one obtains
\be\n{cond}
l^a {\cal A}_a=0\hhh l_a=\nabla_a S\, .
\ee
Similarly, substituting \eq{gop} into \eq{eqn} and keeping the terms
of order $\epsilon^{-2}$ and $\epsilon^{-1}$ one gets
\ba
l_a l^a&=&0\, ,\n{null}\\
l^{b} \nabla_b {\cal A}_{a}&=& -{1\over 2} {\cal A}_a \nabla_b l^{b} \, .\n{polar}
\ea
Since $\nabla_bl_{a}=\nabla_b\nabla_aS=\nabla_a\nabla_bS=\nabla_al_{b}$ the equation \eq{null}
implies that 
\be
l^{b} \nabla_bl^{a}=\nabla_ll^a=0\, .
\ee
Hence  integral lines of $l^a$
\be
{dx^a\over d\tau}=l^{a}
\ee
are null geodesics and $\tau$ is an affine parameter.

We call {\em $l$-plane} a $(D-1)$-dimensional null plane formed by the
vectors $\bs{v}$ orthogonal to $\bs{l}$, $\tens{v}\cdot \tens{l}=0$. 
Relation
\eq{cond} shows that the vector $\tens{{\cal A}}$ lies in the
$l$-plane.
Consider a gauge transformation of the potential $A_a\to
A_a+\nabla_a\alpha$, where $\alpha=\Re \left[\epsilon \gamma
\exp(iS/\epsilon)\right]$. This transformation generates the following map
${\cal A}_a\to {\cal A}_a+\gamma l_a$. This means that the vector
$\tens{{\cal A}}$ is determined up to the transformation 
\be
\tens{{\cal A}}\to \tens{{\cal A}}+{\gamma} \tens{l}\, .
\ee
Thus for a non-trivial electromagnetic field the vector
$\tens{{\cal A}}$ is spacelike. Let us write $\tens{{\cal A}}={\cal A} \tens{e}$, where $\tens{e}\cdot \tens{e}
=1$. We
call ${\cal A}$ the amplitude and $\tens{e}$ the polarization vector.

Since $e^a \nabla_{l} e_a=0$,
the equation \eq{polar} implies
\ba
&&\nabla_{l} \bs{e}=0\, ,\n{pol}\\
&&\nabla_{l} {\cal A}+{1\over 2} {\cal A}\,\nabla_b l^b=0\, .\n{ampl}
\ea
The first equation, \eq{pol}, shows that the vector of polarization
$\bs{e}$ is parallel-transported along the null geodesic, while the
second equations implies 
\be\n{cons}
\nabla_a({\cal A}^2 l^a)=0\, .
\ee
This conserved current gives the conservation law for the `number of
photon'
\be
N=\int d\Sigma_a {\cal A}^2 l^a\, ,
\ee
where $d\Sigma$ is a volume element of a $(D-1)$-dimensional spacelike
Cauchy surface.

Denote by $\bs{e}_{i}$, $i=1, \ldots, D-2$ a set of $(D-2)$
parallel-propagated mutually orthogonal unit vectors. An arbitrary
vector of the linear polarization $\bs{e}$ can be decomposed in this
basis as follows 
\be
\bs{e}=\sum_{i=1}^{D-2} b_{i} \bs{e}_{i}\, ,
\ee
where $b_i$ are constant coefficients.

\section{Parallel transport along principal null directions}
In this appendix we present the details of the construction of a 
parallel-transported frame along the principal null directions.
Besides the basis $\{\tens{l}, \tens{n}, \tens{e}_{\hat \mu},\tens{\tilde e}_{\hat\mu}, \tens{e}_{\hat \epsilon } \}$, it is useful to consider also the complex null Darboux basis 
$\{\tens{l}, \tens{n}, \muv{\mu}, \mbv{\mu}, \tens{e}_{\hat \epsilon } \}$, 
\begin{equation}\label{mdef}
  \muv{\mu}=\srh(\tens{\tilde e}_{\hat \mu}+i\tens{e}_{\hat \mu})\;,\quad\mbv{\mu}=\srh(\tens{\tilde e}_{\hat \mu}-i\tens{e}_{\hat \mu})\;,
\end{equation}
${\mu=1,\dots,n-1}$, with the only non-vanishing scalar products 
\be\label{mm_C}
\tens{l}\cdot\tens{n}=-1\,,\quad \tens{m}_{\hat \mu}\cdot \tens{\bar m}_{\hat \mu}=1\,,
\ee 
in which the PCKY tensor $\tens{h}$, \eqref{KYKNA}, takes the form 
\be\label{he_C}
\tens{h}=r\,\tens{l}^\flat\! \wedge \tens{n}^\flat +i \sum_\mu x_\mu{\tens{m}_\mu}^\flat\wedge {\tens{\bar m}_\mu}^\flat\,.
\ee 

The covariant derivatives of the Darboux basis 
along the principal null direction ${\mom\propto\lv}$ are 
\begin{equation}\label{covdKYfr}
\begin{aligned}
  \nabla_{\!l_{+\!}}\, \lv &= \frac{\sqrt{Q_n}{}_{,r}}{\sqrt{Q_n}}\,\lv\;,\\ 
 \nabla_{\!l_{+\!}}\, \kv &=-\frac{\sqrt{Q_n}{}_{,r}}{\sqrt{Q_n}}\,\kv+\eps\sqrt{2}\,\frac{\sqrt{Q_\epsilon}}{\sqrt{Q_n}}\,\tens{e}_{\hat \epsilon}\\
    & +\sqrt{2}\sum_{\mu=1}^{n-1}\frac{\sqrt{Q_\mu}}{\sqrt{Q_n}}\Bigl(\frac{x_\mu}{x_\mu^2{+}r^2}\,\tens{e}_{\hat \mu}+\frac{r}{x_\mu^2{+}r^2}\,\tens{\tilde e}_{\hat \mu}\Bigr)\;,\\ 
 \nabla_{\!l_{+\!}}\, \tens{e}_{\hat \mu} &= \frac{x_\mu}{x_\mu^2+r^2}\,\tens{\tilde e}_{\hat \mu}+\frac{\sqrt{2}\,x_\mu}{x_\mu^2+r^2}\frac{\sqrt{Q_\mu}}{\sqrt{Q_n}}\,\lv\;,\\ 
  \nabla_{\!l_{+\!}}\, \tens{\tilde e}_{\hat \mu} &= -\frac{x_\mu}{x_\mu^2+r^2}\,\tens{e}_{\hat \mu}+\frac{\sqrt{2}\,r}{x_\mu^2+r^2}\frac{\sqrt{Q_\mu}}{\sqrt{Q_n}}\,\lv\;,\\
 \nabla_{\!l_{+\!}}\, \tens{e}_{\hat \epsilon} &=\frac{\sqrt{2}}{r}\,\frac{\sqrt{Q_\epsilon}}{\sqrt{Q_n}}\,\lv\;.
\end{aligned}
\end{equation}
For the null basis these are equivalent to
\begin{equation}\label{covdKYNfr}
\begin{aligned}
  \nabla_{\!l_{+\!}}\, \muv{\mu} &= \frac{ix_\mu}{x_\mu^2+r^2}\,\muv{\mu}+\frac{i}{x_\mu+ir}\frac{\sqrt{Q_\mu}}{\sqrt{Q_n}}\,\lv\;,\\ 
  \nabla_{\!l_{+\!}}\, \mbv{\mu} &= \frac{-ix_\mu}{x_\mu^2+r^2}\,\mbv{\mu}-\frac{i}{x_\mu-ir}\frac{\sqrt{Q_\mu}}{\sqrt{Q_n}}\,\lv\;,\\
  \nabla_{\!l_{+\!}}\, \kv &=-\frac{\sqrt{Q_n}{}_{,r}}{\sqrt{Q_n}}\,\kv+\eps\sqrt{2}\,\frac{\sqrt{Q_\epsilon}}{\sqrt{Q_n}}\,\tens{e}_{\hat \epsilon}\\
    & +\sum_{\mu=1}^{n-1}\frac{\sqrt{Q_\mu}}{\sqrt{Q_n}}\Bigl(\frac{-i}{x_\mu{-}ir}\,\muv{\mu}+\frac{i}{x_\mu{+}ir}\,\mbv{\mu}\Bigr)\;,
\end{aligned}
\end{equation}
with the equations for ${\lv}$ and ${\tens{e}_{\hat \epsilon}}$ unchanged.

Now, we change these Darboux bases to the parallel-transported ones by a sequence of
the local Lorentz transformations. (Such transformations preserve the orthogonality and the normalization of the frame.) 
Guided by Eq. \eqref{momcoor}, our first transformation is the boost in the $\{\lv, \kv\}$ plane
\begin{equation}\label{boost}
\begin{gathered}
  \bfr\lv = \frac1{\sqrt{Q_n}}\,\lv\;,\quad
  \bfr\kv = \sqrt{Q_n}\,\kv\;,\\
  \bfr\tens{e}_{\hat \mu}=\tens{e}_{\hat \mu}\;,\quad
  \bfr\tens{\tilde e}_{\hat \mu}=\tens{\tilde e}_{\hat \mu}\;,\quad
  \bfr\tens{e}_{\hat \epsilon}=\tens{e}_{\hat \epsilon}\;,\\
  \bfr\muv{\mu}=\muv{\mu}\;,\quad
  \bfr\mbv{\mu}=\mbv{\mu}\;.
\end{gathered}
\end{equation}
The covariant derivatives along ${\mom}$ changes to
\begin{equation}\label{covdBfr}
\begin{aligned}
  \nabla_{\!l_{+\!}}\, \bfr\lv &= 0\;,\\ 
\nabla_{\!l_{+\!}}\, \bfr\kv &= \sqrt{2}\sum_{\mu=1}^{n-1}\sqrt{Q_\mu}\Bigl(\frac{x_\mu}{x_\mu^2{+}r^2}\,\bfr\tens{e}_{\hat \mu}+\frac{r}{x_\mu^2{+}r^2}\,\bfr\tens{\tilde e}_{\hat \mu}\Bigr)\\
      &\quad+\eps\sqrt{2}\,\sqrt{Q_\epsilon}\,\bfr\tens{e}_{\hat \epsilon}\;,\\  
\nabla_{\!l_{+\!}}\, \bfr\tens{e}_{\hat \mu} &= \frac{x_\mu}{x_\mu^2+r^2}\,\bfr\tens{\tilde e}_{\hat \mu}+\frac{\sqrt{2}\,x_\mu\sqrt{Q_\mu}}{x_\mu^2+r^2}\,\bfr\lv\;,\\ 
  \nabla_{\!l_{+\!}}\, \bfr\tens{\tilde e}_{\hat \mu} &= -\frac{x_\mu}{x_\mu^2+r^2}\,\bfr\tens{e}_{\hat \mu}+\frac{\sqrt{2}\,r\sqrt{Q_\mu}}{x_\mu^2+r^2}\,\bfr\lv\;,\\
  \nabla_{\!l_{+\!}}\, \bfr\tens{e}_{\hat \epsilon} &=\sqrt{2}\,\frac{\sqrt{Q_\epsilon}}{r}\,\bfr\lv\;,
\end{aligned}
\end{equation}
or, for the null frame,
\begin{equation}\label{covdBNfr}
\begin{aligned}
  \nabla_{\!l_{+\!}}\, \bfr\muv{\mu} &= \frac{ix_\mu}{x_\mu^2+r^2}\,\bfr\muv{\mu}+\frac{i\sqrt{Q_\mu}}{x_\mu+ir}\,\bfr\lv\;,\\ 
  \nabla_{\!l_{+\!}}\, \bfr\mbv{\mu} &= \frac{-ix_\mu}{x_\mu^2+r^2}\,\bfr\mbv{\mu}-\frac{i\sqrt{Q_\mu}}{x_\mu-ir}\,\bfr\lv\;,\\
  \nabla_{\!l_{+\!}}\, \bfr\kv &= \sum_{\mu=1}^{n-1}\sqrt{Q_\mu}\Bigl(\frac{-i}{x_\mu{-}ir}\,\bfr\muv{\mu}+\frac{i}{x_\mu{+}ir}\,\bfr\mbv{\mu}\Bigr)\\
      &\quad+\eps\sqrt{2}\,\sqrt{Q_\epsilon}\,\bfr\tens{e}_{\hat \epsilon}\;.
\end{aligned}
\end{equation}

Next transformation is a multi-null rotation, leaving ${\bfr\lv}$ fixed. Actually, this transformation can be decomposed into a sequence of commuting null rotations each of which combines only vectors ${\lv,\kv}$ and vectors ${\muv{\mu},\mbv{\mu}}$ from one KY 2-plane. Namely, for each ${\mu=1,\dots,n{-}1}$
we perform the null rotation characterized by the parameter ${\sqrt{Q_\mu}}$\,,
\begin{equation}\label{nullrotm}
\begin{gathered}
  \nfr\lv=\bfr\lv\;,\quad\\
  \nfr\muv{\mu}=\bfr\muv{\mu}+\sqrt{\smash[b]{Q_\mu}}\,\bfr\lv\;,\quad
  \nfr\mbv{\mu}=\bfr\mbv{\mu}+\sqrt{\smash[b]{Q_\mu}}\,\bfr\lv\;,\\
  \nfr\tens{e}_{\hat \mu}=\bfr\tens{e}_{\hat \mu}\;,\quad
  \nfr\tens{\tilde e}_{\hat \mu}=\bfr\tens{\tilde e}_{\hat \mu}+\sqrt{2}\sqrt{\smash[b]{Q_\mu}}\,\bfr\lv\;,
\end{gathered}
\end{equation}
in odd dimension accompanied by
\begin{equation}\label{nullrotz}
  \nfr\tens{e}_{\hat \epsilon}=\bfr\tens{e}_{\hat \epsilon}+\sqrt{2}\sqrt{\smash[b]{Q_\epsilon}}\,\bfr\lv\;.
\end{equation}
The transformed vectors are orthogonal and have to be completed by 
properly transformed vector ${\bfr\kv}$
\begin{equation}\label{nullrotk}
\begin{split}
  \nfr\kv&=\bfr\kv+\sum_{\mu=1}^{n-1}\sqrt{\smash[b]{Q_\mu}}\bigl(\bfr\muv{\mu}{+}\bfr\mbv{\mu}\bigr)+\eps\sqrt{2}\sqrt{\smash[b]{Q_\epsilon}}\,\bfr\tens{e}_{\hat \epsilon}\\
         &\quad\quad\;+\Bigl(\sum_{\mu=1}^{n-1}Q_\mu+\eps\, Q_\epsilon\Bigr)\,\bfr\lv\\
         &=\bfr\kv+\sum_{\mu=1}^{n-1}\sqrt{2}\sqrt{\smash[b]{Q_\mu}}\,\bfr\tens{\tilde e}_{\hat \mu}+\eps\sqrt{2}\sqrt{\smash[b]{Q_\epsilon}}\,\bfr\tens{e}_{\hat \epsilon}\\
         &\quad\quad\;+\Bigl(\sum_{\mu=1}^{n-1}Q_\mu+\eps\, Q_\epsilon\Bigr)\,\bfr\lv\;.
\end{split}
\end{equation}
Let us remark here, that since the parameters of the null rotations are real, 
each of them actually mix only three directions ${\bfr\lv,\bfr\kv,\bfr\tens{\tilde e}_{\hat \mu}}$; the direction ${\bfr\tens{e}_{\hat \mu}}$ remains fixed.
The covariant derivatives of the null rotated frame are simple
\begin{equation}\label{covdnfr}
\begin{gathered}
  \nabla_{\!l_{+\!}}\, \nfr\lv = 0\;,\quad
  \nabla_{\!l_{+\!}}\, \nfr\kv = 0\;,\quad
  \nabla_{\!l_{+\!}}\, \nfr\tens{e}_{\hat \epsilon} = 0\;,\quad\\
  \nabla_{\!l_{+\!}}\, \bfr\tens{e}_{\hat \mu} {=} \frac{x_\mu}{x_\mu^2+r^2}\,\nfr\tens{\tilde e}_{\hat \mu}\;,\quad
  \nabla_{\!l_{+\!}}\, \bfr\tens{\tilde e}_{\hat \mu} {=} -\frac{x_\mu}{x_\mu^2+r^2}\,\nfr\tens{e}_{\hat \mu}\;,\\
  \nabla_{\!l_{+\!}}\,\! \bfr\mbv{\mu} {=} \frac{-ix_\mu}{x_\mu^2+r^2}\,\nfr\mbv{\mu}\;,\;
  \nabla_{\!l_{+\!}}\,\! \bfr\muv{\mu} {=} \frac{ix_\mu}{x_\mu^2+r^2}\,\nfr\muv{\mu}\;.
\end{gathered}
\end{equation}

Finally, we perform the spatial rotation in each spatial KY 2-plane, by the angle ${\ph_\mu=\arctan\frac{r}{x_\mu\!}}$\,,
\begin{equation}\label{srot}
\begin{gathered}
  \pfr\lv=\nfr\lv\;,\quad
  \pfr\kv=\nfr\kv\;,\quad
  \pfr\tens{e}_{\hat \epsilon}=\nfr\tens{e}_{\hat \epsilon}\;,\\
\begin{aligned}
  \pfr\tens{e}_{\hat \mu}&=\frac{x_\mu}{\sqrt{\smash[b]{x_\mu^2{+}r^2}}}\,\nfr\tens{e}_{\hat \mu}-\frac{r}{\sqrt{\smash[b]{x_\mu^2{+}r^2}}}\,\nfr\tens{\tilde e}_{\hat \mu}\;,\\[0.5ex]
  \pfr\tens{\tilde e}_{\hat \mu}&=\frac{r}{\sqrt{\smash[b]{x_\mu^2{+}r^2}}}\,\nfr\tens{e}_{\hat \mu}+\frac{x_\mu}{\sqrt{\smash[b]{x_\mu^2{+}r^2}}}\,\nfr\tens{\tilde e}_{\hat \mu}\;,
\end{aligned}\\[1ex]
  \pfr\muv{\mu}=\sqrt{\frac{x_\mu{-}ir}{x_\mu{+}ir}}\,\nfr\muv{\mu}\;,\quad
  \pfr\mbv{\mu}=\sqrt{\frac{x_\mu{+}ir}{x_\mu{-}ir}}\,\nfr\mbv{\mu}\;.
\end{gathered}
\end{equation}
The resulting frame $\{\pfr\lv,\pfr\kv,\pfr\tens{e}_{\hat \mu},\pfr\tens{\tilde e}_{\hat \mu},\pfr\tens{e}_{\hat \epsilon}\}$
(or, alternatively, $\{\pfr\lv,\pfr\kv,\pfr\muv{\mu},\pfr\mbv{\mu},\pfr\tens{e}_{\hat \epsilon}\}$) is parallel-transported
\begin{equation}\label{covdpfr}
\begin{gathered}
  \nabla_{\!l_{+\!}}\, \pfr\lv = 0\;,\quad
  \nabla_{\!l_{+\!}}\, \pfr\kv = 0\;,\quad
  \nabla_{\!l_{+\!}}\, \pfr\tens{e}_{\hat \epsilon} = 0\;,\quad\\
  \nabla_{\!l_{+\!}}\, \pfr\tens{e}_{\hat \mu} = 0\;,\quad
  \nabla_{\!l_{+\!}}\, \pfr\tens{\tilde e}_{\hat \mu} = 0\;,\\
  \nabla_{\!l_{+\!}}\, \pfr\muv{\mu} = 0\;,\quad
  \nabla_{\!l_{+\!}}\, \pfr\mbv{\mu} = 0\;.
\end{gathered}
\end{equation}

Combining all the transformations together we arrive at the result 
\eqref{partr-KY}, or, for the complex null frame,
\begin{equation}\label{Npartr-KY}
\begin{aligned}
  \pfr\lv&=\frac1{\sqrt{\smash[b]{Q_n}}}\,\lv\;,\\
 \pfr\kv&=\sqrt{\smash[b]{Q_n}}\,\kv+\Bigl(\sum_{\mu=1}^{n-1}Q_\mu+\eps\, Q_\epsilon\Bigr)\frac1{\sqrt{\smash[b]{Q_n}}}\;\lv\;\\
         &\quad+\sum_{\mu=1}^{n-1}\sqrt{\smash[b]{Q_\mu}}\bigl(\muv{\mu}{+}\mbv{\mu}\bigr)+\eps\sqrt{2}\sqrt{\smash[b]{Q_\epsilon}}\;\tens{e}_{\hat \epsilon}\\
  \pfr\muv{\mu}&=\sqrt{\frac{x_\mu{-}ir}{x_\mu{+}ir}}\,\biggl(\muv{\mu}+\frac{\sqrt{\smash[b]{Q_\mu}}}{\sqrt{\smash[b]{Q_n}}}\,\lv\biggr)\;\,\\
  \pfr\mbv{\mu}&=\sqrt{\frac{x_\mu{+}ir}{x_\mu{-}ir}}\,\biggl(\mbv{\mu}+\frac{\sqrt{\smash[b]{Q_\mu}}}{\sqrt{\smash[b]{Q_n}}}\,\lv\biggr)\;\,\\
  \pfr\tens{e}_{\hat \epsilon}&=\tens{e}_{\hat \epsilon}+\sqrt{2}\,\frac{\sqrt{\smash[b]{Q_\epsilon}}}{\sqrt{\smash[b]{Q_n}}}\,\lv\;.
\end{aligned}
\end{equation}

\section{Parallel transport in the Pleba\'nski-Demia\'nski family of solutions}   
In the main text we have demonstrated how to construct a parallel-propagated frame along null geodesics in spacetimes admitting the PCKY tensor.
In this appendix we show how to modify this construction for an important family of 4D spacetimes described by the Pleba\'nski-Demia\'nski metric.
Such a family generally admits only a non-closed generalization of the PCKY tensor---the (non-degenerate) conformal Killing--Yano (CKY) tensor.
Our construction generalizes the results presented in \cite{Marck:1983null,KamranMarck:1986}.

\subsection{Pleba\'nski-Demia\'nski metric}
The Pleba\'nski-Demia\'nski metric \cite{PlebanskiDemianski:1976} describes a large class of four-dimensional type D spacetimes. The concrete form of physical metrics is obtained by a due limiting procedure (see, e.g., \cite{GriffithsPodolsky:2006b} for a recent review). In this  way one can obtain, for example, the metric of an accelerated rotating charged black hole in the cosmological background. 

The whole Pleba\'nski-Demia\'nski class of solutions possesses the CKY tensor \cite{KubiznakKrtous:2007}. 
Such a tensor is responsible for complete integrability of a null geodesic motion.
When the acceleration parameter is removed the corresponding subclass of solutions obtained earlier by Carter \cite{Carter:1968cmp, Carter:1968pl} allows the PCKY tensor \cite{Carter:1987} and the solution of parallel transport was already described in Section VI.A (see also \cite{ConnellEtal:2008} for the
timelike case). 
In order to see the impact of the presence of a non-trivial acceleration on  parallel transport, we write the  Pleba\'nski-Demia\'nski metric in the notations of Section VI.A.
So we have
\be\label{PD}
\tens{g}=-\tens{\tilde \omega}^{\hat 2}\tens{\tilde \omega}^{\hat 2}+\tens{\omega}^{\hat 2}\tens{\omega}^{\hat 2}+
\tens{\tilde \omega}^{\hat 1}\tens{\tilde \omega}^{\hat 1}+\tens{\omega}^{\hat 1}\tens{\omega}^{\hat 1}\,,
\ee
where  
\begin{equation}\label{tetradPD}
\begin{split}
\tens{\tilde \omega}^{\hat 2}=&\,\Omega\sqrt{\frac{X_2}{U_2}}(\tens{d}\psi_0+x_1^2\tens{d}\psi_1 )\,,\ 
\tens{\omega}^{\hat 2}=\,\Omega\sqrt{\frac{U_2}{X_2}}\,\tens{d}r\,,\\ 
\tens{\tilde \omega}^{\hat 1}=&\,\Omega\sqrt{\frac{X_1}{U_1}}(\tens{d}\psi_0-r^2\tens{d}\psi_1 )\,,\ 
\tens{\omega}^{\hat 1}=\Omega\sqrt{\frac{U_1}{X_1}}\, \tens{d}x_1\,. 
\end{split}
\end{equation}
Here, $U_2=-U_1=x_1^2+r^2$ and $\Omega=(1-x_1r)^{-1}$. 
For $X_1=X_1(x_1)$ and $X_2=X_2(r)$  we refer to the metric as the {\em off-shell} metric. 
Such a metric possesses two (Hodge dual) non-degenerate CKY 2-forms
\cite{KubiznakKrtous:2007}
\ba
\tens{h}\!\!&=&\!\!\Omega\left(-r\tens{\omega}^{\hat 2}\wedge\tens{\tilde \omega}^{\hat 2}+x_1\tens{\omega}^{\hat 1}\wedge\tens{\tilde \omega}^{\hat 1}\right)\,,\\
\tens{k}\!\!&=&\!-\Omega\left(x_1\tens{\omega}^{\hat 2}\wedge\tens{\tilde \omega}^{\hat 2}+r\tens{\omega}^{\hat 1}\wedge\tens{\tilde \omega}^{\hat 1}\right)\,,
\ea
which are connected with the background isometries as
\ba\label{KVS}
\tens{\xi}_{(h)}\!\!&=&\!\!-\frac{1}{3}\tens{\delta h}=\pa_{\psi_0}\,,\nonumber\\
\tens{\xi}_{(k)}\!\!&=&\!\!-\frac{1}{3}\tens{\delta k}=\pa_{\psi_1}\,.
\ea
These isometries together with the hidden symmetry of the CKY tensor make the null geodesic motion in the off-shell background \eqref{PD}, \eqref{tetradPD} completely integrable. The tetrad components of the velocity are 
\be
\begin{split}
{\tilde l}_{\hat 2}=&\,\frac{W_2}{\Omega\sqrt{X_2U_2}}\,,\  
l_{\hat 2}=\frac{\sigma_2}{\Omega\sqrt{X_2U_2}}\sqrt{W_2^2-V_2X_2}\,,\\  
{\tilde l}_{\hat 1}=&\,\frac{-W_1}{\Omega\sqrt{X_1U_1}}\,,\  
l_{\hat 1}=\frac{\sigma_1}{\Omega\sqrt{X_1U_2}}\sqrt{V_1X_1-W_1^2}\,, 
\end{split}
\ee
where
\begin{equation}
\begin{split}
W_2=&\,r^2\Psi_0+\Psi_1\,,\quad V_2=-\kappa_1>0\,,\\
W_1=&\,-x_1^2\Psi_0+\Psi_1\,,\quad V_1=\kappa_1\,.
\end{split}
\end{equation}
The constant $\kappa_1$ corresponds to the conformal Killing tensor ${Q}_{ab}=k_{ac}k_b^{\ c}$, whereas the constants $\Psi_0$ and $\Psi_1$ are associated with the Killing vectors $\pa_{\psi_0}$ and $\pa_{\psi_1}$, respectively.\footnote{%
One can formally recover the `non-accelerating' class of solutions \cite{Carter:1968cmp}, \cite{Carter:1968pl} by taking the conformal factor $\Omega=1$. After that, the CKY tensor $\tens{h}$ becomes the PCKY tensor discussed earlier and $\tens{k}$ becomes the Killing--Yano tensor. At the same time $\tens{\xi}_{(h)}$ still coincides with $\pa_{\psi_0}$ and $\tens{\xi}_{(k)}$ vanishes.
}

For the special choice of metric functions
\be
\begin{split}
X_1\!=&-\!k\!-\!2nx_1\!+\!\epsilon x_1^2\!-\!2mx^3_{1}\!+\!(k\!+\!e^2\!+\!g^2\!\!+\!\Lambda/3)x^4_{1}\,,\\
X_2\!=&\,k\!+\!e^2\!+\!g^2\!-\!2mr\!+\!\epsilon r^2\!-\!2nr^3\!-\!(k\!+\!\Lambda/3)r^4\,,
\end{split}
\ee
and the vector potential
\ba
\tens{A}=-\frac{1}{\Omega}\Bigl(\frac{er}{\sqrt{U_2X_2}}\,\tens{\tilde \omega}^{\hat 2}+\frac{gx_1}{\sqrt{U_1X_1}}\,\tens{\tilde \omega}^{\hat 1}\Bigr)\,,
\ea 
the Pleba\'nski-Demia\'nski metric obeys the  Einstein--Maxwell equations with 
$e$ and $g$ the electric and magnetic charges and the cosmological constant $\Lambda$.

\subsection{Construction of parallel-propagated frame}
Let us now construct a parallel-propagated frame in the off-shell background \eqref{PD}, \eqref{tetradPD}.
We start by observing that having a CKY 2-form $\tens{\omega}$, that is a 2-form obeying
the CKY equations
\be\label{CKY2}
\nabla_{X}\tens{\omega}=\frac{1}{3}\tens{X}\hook \tens{d\omega}+\bs{X}^{\flat}\!\wedge \tens{\xi}^\flat\,, \ \tens{\xi}^\flat=-{1\over D-1}\, \tens{\delta \omega},
\ee 
and a null geodesic velocity vector $\tens{l}$, one can construct the following parallel-transported vector:
\begin{equation}\label{mcky}
\tens{w}=(\tens{l}\hook \tens{\omega})^\sharp +\beta\tens{l}\,,\quad 
\dot \beta=\tens{l \cdot \xi}\,.
\end{equation}
Indeed, using the defining property \eqref{CKY2} we obtain
\be\label{proof2}
\begin{split}
\tens{\dot w}&=(\tens{l}\hook \tens{\dot \omega})^\sharp +\dot \beta \tens{l}\\
&=\frac{1}{3}\,\tens{l}\hook(\tens{l}\hook\tens{d\omega})+\bigr[\tens{l}\hook(\tens{l}^\flat\wedge \tens{\xi}^\flat)\bigr]^\sharp +\dot \beta \tens{l}\\
&=\tens{l}(\dot \beta-\tens{l \cdot \xi})\,.
\end{split}
\ee
Here we have used the obvious fact that the first term in the second line is zero and then proceeded in the same way as in Section III.A.

In particular, this means that for the off-shell Pleba\'nski-Demia\'nski metric we may construct the following two parallel-transported 
vectors:
\ba
\tens{m}\!\!&=&\!\!\frac{1}{\sqrt{-\kappa_1}}\bigl[ (\tens{l}\hook \tens{h})^\sharp+\beta_h\tens{l} \bigr]\,,\\
\tens{z}\!\!&=&\!\!\frac{1}{\sqrt{-\kappa_1}}\bigl[ (\tens{l}\hook \tens{k})^\sharp+\beta_k\tens{l} \bigr]\,.
\ea
Moreover, using \eqref{KVS} we find\footnote{%
Let us remark here, that contrary to  Section IV.B one cannot for the 
Pleba\'nski-Demia\'nski metric `separate' the affine parameter $\tau$ as in \eqref{tau}.
Formally, this is due to the presence of a nontrivial (non-separable) conformal factor $\Omega$.}
\be
\beta_h=\Psi_0\tau\,,\quad \beta_k=\Psi_1\tau\,.
\ee
The last parallel-transported vector $\tens{n}$ is simply determined by the normalization conditions
\be
\tens{n \cdot l}=-1\,,\ \ 
\tens{n\cdot m}=0\,,\ \ 
\tens{n\cdot z}=0\,\ \ 
\tens{n \cdot n}=0\,.
\ee 

Let us explicitly write down the form of the (of-shell) parallel-transported frame 
in the velocity adapted basis $\{\tens{o}\}$, \eqref{newframe}. We have
\be
\begin{split}
\tens{l}^\flat\!=&\,{\tilde k}_{\hat 1}(-\tens{\tilde o}^{\hat 2}+
\tens{\tilde o}^{\hat 1})\,,\quad {\tilde k}_{\hat 1}=-{\tilde k}_{\hat 2}=\frac{1}{\Omega}\frac{\sqrt{V_2}}{\sqrt{U_2}}\,,\\
\tens{m}^{\flat}\!=&\,\frac{{\tilde k}_{\hat 1}}{\sqrt{-\kappa_1}}
\bigl(-\beta_h\tens{\tilde o}^{\hat 2}+
r\Omega\tens{ o}^{\hat 2}+\beta_h\tens{\tilde o}^{\hat 1}
-x_1\Omega\tens{ o}^{\hat 1}\bigr)\,,\\
\tens{n}^{\flat}\!=&\,\frac{{\tilde k}_{\hat 1}}{\kappa_1}
\Big[\frac{\Omega^2U_2\!+\!\beta_h^2\!+\!\beta_k^2}{2}\,\tens{\tilde o}^{\hat 2}-\Omega(\beta_h r\!+\!\beta_k x_1)\tens{o}^{\hat 2}\\
\!&+\frac{\Omega^2U_2\!-\!\beta_h^2\!-\!\beta_k^2}{2}\,\tens{\tilde o}^{\hat 1}+\Omega(\beta_h x_1\!-\!\beta_kr)\tens{o}^{\hat 1}\Big]\,,\\
\tens{z}^{\flat}\!=&\,\frac{{\tilde k}_{\hat 1}}{\sqrt{-\kappa_1}}
\bigl(-\beta_k\tens{\tilde o}^{\hat 2}
+x_1\Omega\tens{o}^{\hat 2}
+\beta_k\tens{\tilde o}^{\hat 1}
+r\Omega\tens{o}^{\hat 1}
\bigr)\,.
\end{split}
\ee
It is easy to see, that one can formally recover the `non-accelerating' limit of the PCKY tensor, Eq. \eqref{vysl_4D}, by setting $\Omega=1$ ($\beta_k=0$).

%\medskip
%\vspace{3.5cm}
\section*{Acknowledgments}
\vspace*{-1ex}
D.K. is grateful to the Herchel Smith Postdoctoral Research Fellowship at the University of Cambridge and acknowledges the Golden Bell Jar Graduate
Scholarship in Physics at the University of Alberta. 
V.F. thanks the Natural Sciences and Engineering Research Council of Canada and
the Killam Trust for the financial support.
P.K. is supported by Grant No.~GA\v{C}R 202/06/0041 and the
Czech Ministry of Education under Project No.~MSM0021610860. P.C. is grateful
to the Department of Physics at the University of Alberta for continued financial support.

%\bibliography{Databaze}
%\bibliographystyle{apsrev}

\end{document}